\documentclass[twocolumn,showpacs,preprintnumbers,amsmath,amssymb,prb,superscrip
taddress]{revtex4} 
 
\usepackage{epsfig}
\usepackage[dvips]{color}
\usepackage{graphicx}
\usepackage{dcolumn}
\usepackage{bm}
 
\begin{document} 
 
\title{Multicolored quantum dimer models, resonating valence--bond states, 
color visons, and the triangular--lattice $t_{2g}$ spin--orbital system}
 
\author{B. Normand} 
\affiliation{Department of Physics, Renmin University of China, Zhongguancun 
Ave. 59, Beijing 100872, China} 
 
\date{\today} 

\begin{abstract} 

The spin--orbital model for triply degenerate $t_{2g}$ electrons on a 
triangular lattice has been shown to be dominated by dimers: the phase 
diagram contains both strongly resonating, compound spin--orbital dimer 
states and quasi--static, spin--singlet valence--bond (VB) states. To 
elucidate the nature of the true ground state in these different regimes, 
the model is mapped to a number of quantum dimer models (QDMs), each of 
which has three dimer colors. The generic multicolored QDM, illustrated 
for the two-- and three--color cases, possesses a topological color 
structure, ``color vison'' excitations, and broad regions of resonating 
VB phases. The specific models are analyzed to gain further insight into 
the likely ground states in the superexchange and direct--exchange limits 
of the electronic Hamiltonian, and suggest a strong tendency towards VB 
order in all cases.

\end{abstract} 
 
 
\maketitle 

\section{Introduction} 

Spin--orbital models have become a very active field of research in the 
quest for exotic states of matter and novel properties in systems with 
coupled charge, spin, and orbital degrees of freedom. In real materials,
each class of active orbital, lattice structure, and extent of electron 
filling leads to a different set of problems. An overview of this fabulous 
wealth of possibilities, focusing on undoped systems and including a complete 
review of relevant materials, may be found in Ref.~[\onlinecite{rno}]. 

Among these many models, one of the most exotic and mysterious is 
the triangular--lattice $t_{2g}$ spin--orbital model considered in 
Ref.~[\onlinecite{rno}]. By this is meant the insulating system 
with a single, fully localized electron on each lattice site, and 
occupying one of the threefold--degenerate $t_{2g}$ orbitals, a 
situation which would be realized, for example, in undistorted NaTiO$_2$ 
or, with holes, in CoO$_2$. Depending on which orbitals they occupy, the 
particles have both superexchange and direct--exchange interactions whose 
ratio is {\it a priori} unknown, but it was shown\cite{rno} that different 
types of bond dimer state of the spin and orbital degrees of freedom are 
favored very strongly for all values of this ratio. The resulting, very 
rich, phase diagram of possible states includes candidate resonating 
valence--bond (RVB) states in the limit of pure superexchange interactions 
and, in the direct--exchange limit, candidate systems for extremely subtle 
order--by--disorder selection of special valence--bond crystal (VBC) states 
from very highly degenerate manifolds of dimer coverings. 

The question left unanswered despite the extensive energetic studies 
and plausibility arguments in Ref.~[\onlinecite{rno}] was whether one 
could gain more explicit indications for the suggested unconventional 
nature of the true ground state in either of these cases. The exotic 
phenomena which are known from simplified models of highly frustrated 
and degenerate systems include true liquid phases with no unbroken 
symmetry in the ground state, topological sectors and excitations, 
and fractionalization and deconfinement of elementary spinon--type 
quasiparticles. Given the exhaustive search for both experimental 
and theoretical realizations of such phenomena in real, electronic systems,
the triangular--lattice $t_{2g}$ spin--orbital model is clearly a prime 
candidate for further investigation. Here it can be noted that the 
preponderance of evidence in favor of RVB states suggests further that 
the candidate (spin--orbital) liquid phase is gapped, and hence would 
have only short--ranged correlation functions and massive, fractional 
spinon--orbiton excitations. The aim of this study is to approach the 
unanswered questions through effective quantum dimer models (QDMs). 

A review of QDMs, including their origin, properties, and rich associated 
physics, can be found in Ref.~[\onlinecite{rmr}]. For the present purposes, 
specifically pursuing the tantalizing prospect of an RVB state in a real 
electronic system, the agenda is that laid out as a practical prescription 
in Ref.~[\onlinecite{rmvrbfp}], namely to map the starting Hamiltonian to 
the triangular--lattice QDM. In the original QDM exposition of Rokhsar and 
Kivelson,\cite{rrk} it was shown for the square lattice that an exact RVB 
state exists at one point in the phase diagram (henceforth an ``RK point''), 
and it was later found\cite{rms} that the same model on the triangular 
lattice allows a rigorous proof for the existence of a true RVB phase with 
gapped spinon excitations over a finite regime of parameter space. This 
therefore offers not only a finite chance of any given model having the 
appropriate energetics, but also an approach which captures automatically 
the topological criteria associated with RVB physics.

The generic QDM phase diagram\cite{rmr} nevertheless contains RVB states 
over at best a rather narrow range of parameters. Its bulk is occupied 
by static ``columnar'' and ``staggered'' dimer regimes, which respectively 
maximize or minimize the number of active four--site units, or by other 
possibilities with a more complex breaking of translational symmetry on 
generalized plaquettes. Thus the same approach can also be adopted for 
the direct--exchange limit of the starting model,\cite{rno} not in this 
case with intent to find an RVB phase, but with a view to isolating the 
leading fluctuation term which might be responsible for selecting one, 
or a set of, dimer coverings which are more energetically favorable, 
through the quantum fluctuations they allow, than all others.  

Because QDMs are in general highly simplified, any mapping from a realistic 
Hamiltonian discards of necessity many degrees of freedom, and thus there is 
no systematic procedure for their derivation. When proceeding from real, 
$S = 1/2$, spin--singlet dimers, an immediate dichotomy arises between the 
definition of QDM states as mutually orthogonal and the fact that no 
conventional spin--dimer states are orthogonal (the overlap factor being 
trivially $1/\sqrt{2}$ per site). This non--orthogonality problem is 
strongly reduced in the triangular--lattice $t_{2g}$ spin--orbital model, 
and entirely absent in its direct--exchange limit, for reasons which will 
become apparent in Sec.~II.  However, this fact also forbids the type of 
dimer overlap expansion exploited in Ref.~[\onlinecite{rrk}], and mandates 
instead a process which proceeds directly from the electronic Hamiltonian.
The process of deriving an effective QDM from an orbitally degenerate 
electronic Hamiltonian was followed in Ref.~[\onlinecite{rvrbm}], in the 
course of an analysis of LiNiO$_2$ (a system of $e_g$ electrons on a 
triangular lattice), and while the rationale is similar here, the 
orthogonality of dimer coverings requires an approach somewhat different 
in detail. 

As a foretaste of the results to follow, in the superexchange limit one 
finds two different types of ``three--color'' QDM. The individual dimers 
are either spin singlets with three different flavors corresponding to 
the triplet states of the orbitals (ss/ot), or orbital singlets with a 
conventional spin--triplet flavor (os/st). In both cases, the 9$\times$9 
$t$ and $v$ matrices of the QDM reflect the strong breaking of translational 
and rotational symmetry associated with the consideration of four--site 
plaquette units in the definition of the QDM. In the former case this 
drives a lifting of degeneracy in favor of the $T_z = \pm 1$ components 
of the orbital triplet, while in the latter the spin--triplet dimers may 
continue to interchange their flavors. 

In the direct--exchange limit, there is no mixing between two pairs of 
dimers on a plaquette, and both the $t$ and $v$ terms of the conventional 
QDM are identically zero; this rigid locking of dimer color to lattice 
direction means that the effective model reduces to the one--color variant. 
The relevant QDM capturing the lowest--order dimer resonance processes is 
in fact one defined on six--bond triangular loops. While this model dictates 
a specific ground state, selected from the extensively degenerate manifold 
of dimer coverings, it does not exclude the possibility of a type of 
one--dimensional physics\cite{rno} where the character of short, fluctuating 
segments of frustration--decoupled spin chains persists despite the high site
coordination of the triangular lattice. 

Finally, it should be stressed that none of the physics discussed 
here is to be confused with that of the ``spin--orbital liquid'' 
and ``spin--orbital singlet'' states introduced recently in 
Ref.~[\onlinecite{rcbs}]. The work of these authors depends intrinsically 
on a strong spin--orbit coupling interaction $\lambda_0 {\vec L} \cdot 
{\vec S}$, and should perhaps therefore be referred to more strictly as 
``(spin-orbit)al'' physics. In the present study, which follows a line 
dating back to the seminal work of Kugel' and Khomskii,\cite{rkk} the 
terminology ``spin--orbital'' refers only to the connection, dictated 
precisely by the electronic Hamiltonian, between the magnetic exchange 
interactions and the orbital state of each ion, with $\lambda_0$ assumed 
to be negligible. While this approximation is usually taken to be 
reasonable for $3d$ ions, $\lambda_0$ may indeed become important when 
the physics of the system depends sensitively on the lifting of high 
degeneracies. 

The structure of this manuscript is as follows. The triangular--lattice 
$t_{2g}$ spin--orbital model is presented in Sec.~II, in the form of a 
minimal review of the contents of Ref.~[\onlinecite{rno}] required as a 
basis for the current analysis. The orthogonality of electronic states 
with different spin or orbital colors (colored dimers) is demonstrated 
and the derivation of appropriate QDMs outlined. In Sec.~III, the generic 
multicolored QDM is defined and its properties are studied, illustrating 
the circumstances under which the color degree of freedom leads to 
additional topological sectors, possibilities for RK points and RVB ground 
states, and ``color vison'' excitations associated with the topological 
properties. Section IV returns to a detailed analysis of the specific 
QDMs deduced for the three cases of most interest here, namely (ss/ot) 
and (os/st) dimers in the superexchange limit and spin--singlet dimers 
in the direct--exchange limit. A short discussion and summary are 
presented in Sec.~V. 

\begin{figure}[t!]
\begin{center}
\includegraphics[width=7.5cm]{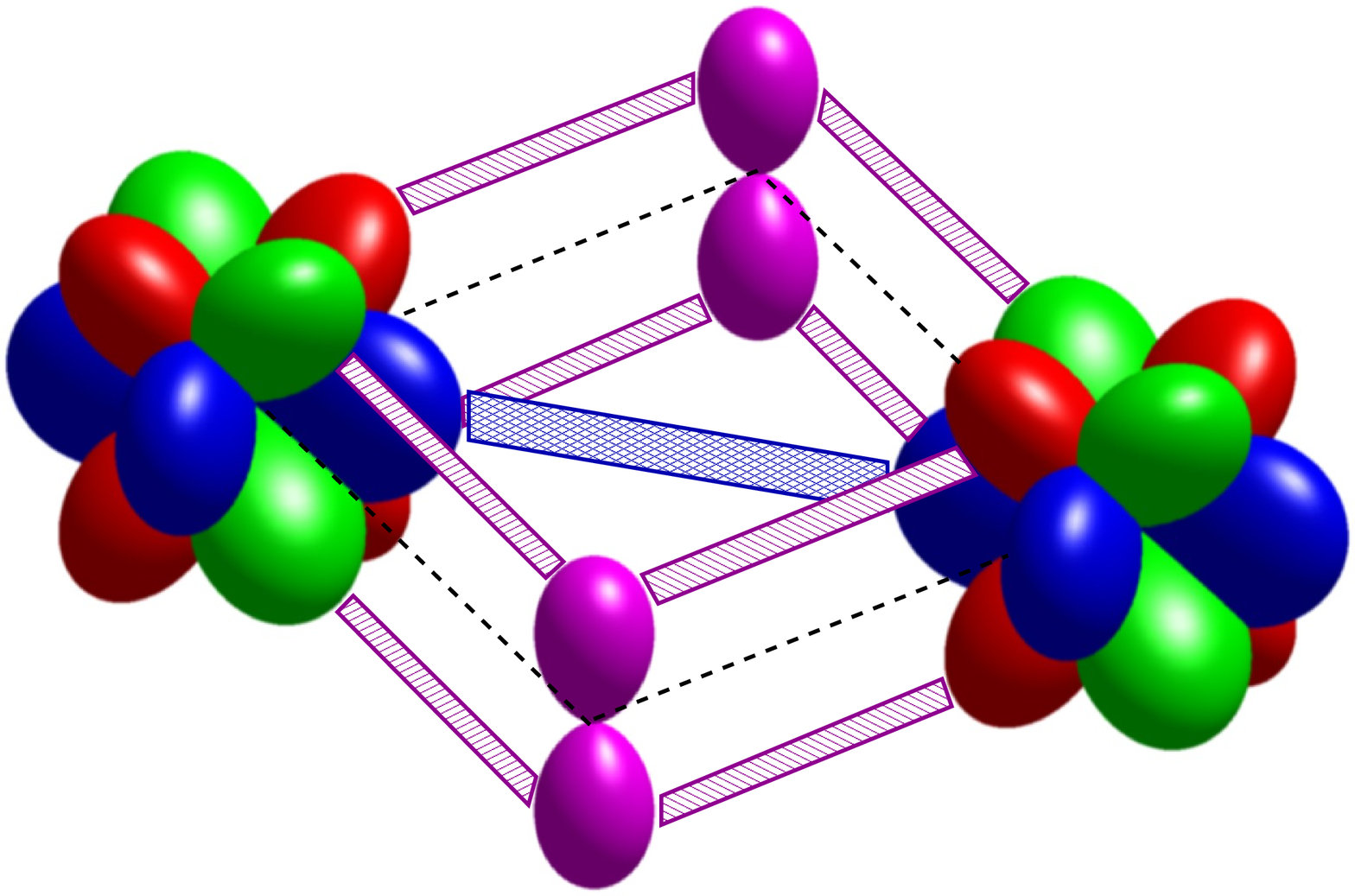}
\centerline{(a)} \vskip .1cm
\includegraphics[width=7.7cm]{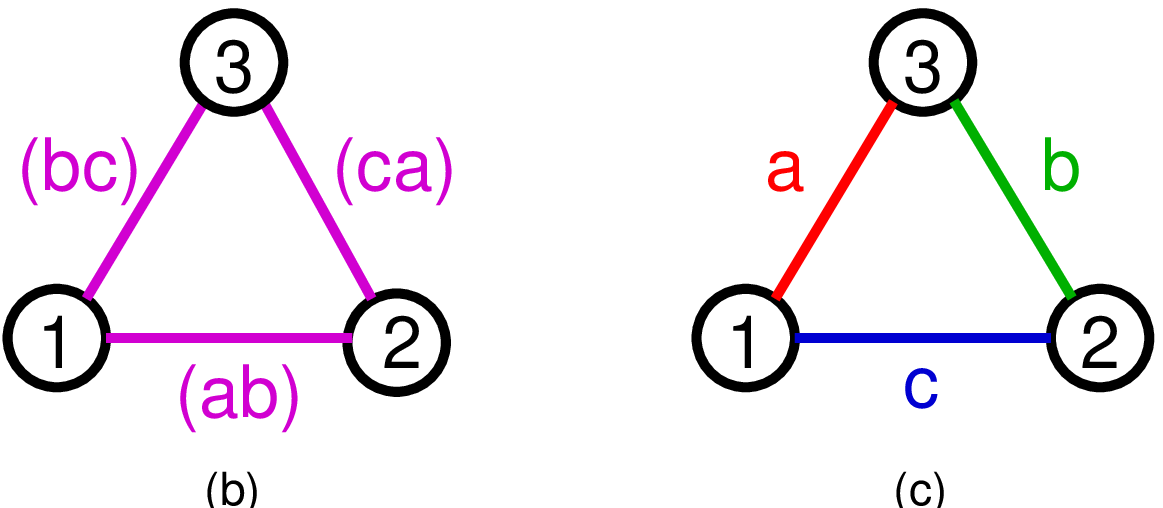}
\end{center}
\caption{(color online) (a) Schematic representation of hopping processes 
for $t_{2g}$ electrons in triangular geometry which contribute to magnetic
interactions on a representative ($c$--axis) bond $\langle ij \rangle$. 
The orbitals are represented by different colors (greyscale intensities).
Superexchange processes involve O $2p_z$ orbitals (violet), and couple 
pairs of $a$ and $b$ orbitals (red, green) with effective hopping elements 
$t_e$, interchanging their orbital color. Direct exchange couples $c$ orbitals 
(blue) with hopping strength $t_e'$. (b) Pairs of $t_{2g}$ orbitals active in 
superexchange and (c) single orbitals active in direct exchange; horizontal 
bonds correspond to the situation depicted in panel (a). }
\label{fig1}
\end{figure}

\section{Model}

The derivation begins with the Hubbard Hamiltonian for $t_{2g}$ electrons 
hopping on a triangular lattice, usually realized on the $\langle 111 
\rangle$ planes of a cubic, perovskite--based structure whose local 
symmetry preserves the orbital degeneracy. The geometry of the system 
is summarized in Fig.~1, and the average filling of 1 electron per site 
corresponds to Ti$^{3+}$ or V$^{4+}$ ions. To avoid undue repetition of 
published material, only the minimal numbers of equations and explanations 
required for a self--consistent presentation are included here; the reader 
wishing a complete exposition is requested to consult Ref.~[\onlinecite{rno}]. 

Specializing immediately to a magnetic insulator with on--site repulsion $U$ 
much greater than the diagonal and off--diagonal (in orbital ``color'') 
electronic hopping integrals, respectively $t_e'$ and $t_e$, the site 
occupation is strictly $n_{ia} + n_{ib} + n_{ic} = 1$. The orbitals have a 
direct connection to the specific lattice directions, and in Fig.~1 the color 
red for the $d_{yz}$ orbital, is associated with $a$, green ($d_{xz}$) 
with $b$, and blue ($d_{xy}$) with $c$. At second order in a perturbative 
treatment, the magnetic Hamiltonian can be written in the schematic form 
\begin{equation}
\label{emh}
{\cal H} = J_s {\cal H}_s + J_m {\cal H}_m + J_d {\cal H}_d,
\end{equation}
where 
\begin{equation}
\label{ej}
J_s = \frac{4t_e^2}{U},    \hskip .5cm
J_m = \frac{4t_et_e'}{U},    \hskip .5cm
J_d = \frac{4t_e'^2}{U}. 
\end{equation}

Superexchange contributions to ${\cal H}$ can be expressed in the form
\begin{eqnarray}
\label{ehs}
{\cal H}_s \! & = & \! \frac{1}{2} \sum_{\langle ij \rangle \parallel \gamma}
\Big\{ r_1\Big( \vec{S}_i \! \cdot \! \vec{S}_j + \frac{3}{4} \Big) \Big[ 
A_{ij}^{\gamma} \! + \frac{1}{2} (n_{i\gamma} + n_{j\gamma}) - 1 \Big] 
\nonumber \\ & & \;\;\;\;\;\; + r_2 \Big( \vec{S}_i \! \cdot \! \vec{S}_j
 - \frac{1}{4}\Big) \Big[ A_{ij}^{\gamma} \! - \frac{1}{2} (n_{i\gamma}
 + n_{j\gamma}) + 1 \Big] \nonumber \\ & & \;\;\;\;\;\; - \frac{2}{3} 
(r_2 - r_3) \Big( \vec{S}_i \! \cdot \! \vec{S}_j - \frac{1}{4} \Big) 
B_{ij}^{\gamma} \Big\},
\end{eqnarray}
where the coefficients 
\begin{equation}
\label{er}
r_1 = \frac{1}{1 - 3 \eta},  \hskip 0.5cm
r_2 = \frac{1}{1 - \eta},    \hskip 0.5cm
r_3 = \frac{1}{1 + 2 \eta},
\end{equation}
are dictated by the different possible energy states of the virtual 
$d_i^1 d_j^1 \rightleftharpoons d_i^2 d_j^0$ excitation, which are 
determined by the Hund coupling $J_H = \eta U$. In real $3d$ 
transition--metal ions, $\eta$ is of order 0.10--0.15. Under a local 
transformation which takes account of the color--exchanging nature 
of the superexchange hopping processes, 
\begin{eqnarray}
\label{epsspd}
\hskip -.7cm
A_{ij}^{\gamma} \!\! & = & \!\! 2 \Big( {\vec T}_{i\gamma} \cdot {\vec
T}_{j\gamma} \! + \! \frac{1}{4} n_i^{\gamma} n_j^{\gamma} \Big),\\
\hskip -.7cm
B_{ij}^{\gamma} \! & = & \!\! 2 \Big( \vec{T}_{i\gamma} \times
\vec{T}_{j\gamma} + \frac{1}{4} n_i^{\gamma} n_j^{\gamma} \Big),
\end{eqnarray}
where the scalar product in $A_{ij}$ is the conventional expression for 
pseudospin--1/2 variables and the cross product in $B_{ij}$ is defined as
\begin{equation}
\vec{T}_{i\gamma} \times \vec{T}_{j\gamma} = \frac12 (
T_{i\gamma}^ + T_{j\gamma}^+ + T_{i\gamma}^- T_{j\gamma}^- ) +
T_{i\gamma}^z T_{j\gamma}^z.
\end{equation}
Contributions of the form $B_{ij}$ vanish as $\eta \rightarrow 0$ 
(\ref{ehs}). In these expressions, $n_i^{\gamma}$ denotes the number 
of superexchange--active electrons (those able to hop on the bond in 
question) and $n_{i\gamma}$ the number of superexchange--inactive 
electrons (which, however, are active in direct--exchange processes): 
thus the pseudospin scalar and vector product interactions are relevant 
only when the electrons on both sites are active, while situations with 
only one active orbital still give nontrivial, spin--dependent 
contributions.

It is clear that the superexchange contribution on a single bond is minimized 
either by an orbital--singlet, spin--triplet state, or by a spin--singlet, 
orbital--triplet state [henceforth (os/st) and (ss/ot)]. Their energies 
\begin{eqnarray}
\label{eosst}
   E_{\rm (os/st)} & = & - J r_1,            \\
\label{essot}
   E_{\rm (ss/ot)} & = & - \frac{1}{3} J \left( 2 r_2 + r_3 \right),
\end{eqnarray}
are degenerate
for $\eta = 0$, while the (os/st) state is favored for finite $\eta$. 
Because the superexchange hopping term is off--diagonal, the orbital 
singlet is the state
\begin{equation}
|\psi_{os} \rangle = \frac{1}{\sqrt{2}} \left( |aa \rangle - |bb
\rangle \right),
\end{equation}
in the original electronic basis, while the orbital triplets are
\begin{eqnarray}
\label{eotp}
|\psi_{ot+} \rangle & = & |ab \rangle,                           \\
\label{eot0}
|\psi_{ot0} \rangle & = & \frac{1}{\sqrt{2}}
                     \left( |aa \rangle + |bb \rangle\right),    \\
\label{eotm}
|\psi_{ot-} \rangle & = & |ba \rangle.
\end{eqnarray}

Contributions from direct--exchange processes take the form 
\begin{eqnarray}
\label{ehd}
{\cal H}_d \! & = & \! \frac{1}{4} \sum_{\langle ij \rangle \parallel \gamma}
\Big\{ \Big[ - r_1 \Big( \vec{S}_i \! \cdot \! \vec{S}_j + \frac{3}{4}\Big)
 + r_2 \Big( \vec{S}_i \! \cdot \! \vec{S}_j - \frac{1}{4} \Big) \Big]
\nonumber \\ & & \hskip 1.2cm \times \Big[ n_{i\gamma} (1 - n_{j\gamma})
 + (1 - n_{i\gamma}) n_{j\gamma} \Big] \\ & & \hskip 1.2cm + \frac{1}{3}
\left( 2r_2 + r_3 \right) \Big( \vec{S}_i \! \cdot \! \vec{S}_j - \frac{1}{4}
\Big) \; 4 n_{i\gamma} n_{j\gamma} \Big\}, \nonumber
\end{eqnarray}
where it is clear once again that far the most favorable energies are 
obtained from dimer states, but only those creating spin singlets from 
two electrons with the bond color. 
Expressions for $H_m$ can be found in Ref.~[\onlinecite{rno}], but, as 
explained there, the very different nature of the two types of singlet 
state mean that $H_m$ has a qualitative effect on the physics of the system 
only in rare situations.

Before analyzing the model in more detail, it is necessary to delimit the 
parameter space of interest to this study, and also to make one essential 
and completely general comment concerning dimer states of colored electrons. 
First, the focus of this manuscript is only the two limits of pure 
superexchange and purely direct exchange. In the superexchange limit, only 
the regimes with Hund coupling ratio $\eta = 0$ and with physical values 
$\eta \sim 0.1$ (which is ``large'' in a sense to be made explicit below), 
will be considered. In the direct--exchange limit, only $\eta = 0$ is of 
interest in the context of QDMs: the physics of dimer--based models is 
determined by purely static contributions at any finite $\eta$,\cite{rji} 
and changes away from dimer--based states only at the unrealistically large 
value $\eta = 0.2$. Thus there are three situations to elucidate, all of 
which have already been shown to be dominated by dimer formation. This 
restriction is made partly for reasons of practicality, partly to concentrate 
on the essentially different physics of the differing limits, and partly as 
a consequence of the above observation concerning the irrelevance of $H_m$ 
and absence of meaningful mixed processes. 

\subsection{Effective dimer overlap}

As noted above, a serious problem in mapping real spin models to QDMs is 
the ``non-orthogonality catastrophe,'' the fact that all dimer coverings 
$|c \rangle$ have finite overlap with all others. This is a significant 
impediment to many forms of both analytical and numerical progress. For 
$S = 1/2$ entities forming bond spin--singlet states, the generic overlap 
of two singlets at the same site is simply the prefactor of the singlet 
wave function, $\alpha = 1/\sqrt{2}$. In the present context, the most 
meaningful calculation involving $\alpha$ is to consider the overlap 
of the wave functions of two ``horizontal'' and two ``vertical'' SU(2) 
spin singlets on a four--site plaquette, where it is easy to show from 
the common elements of $|h \rangle = \alpha^2 (|1 \!\! \uparrow 2 \!\!\! 
\downarrow \rangle - |1 \!\!\! \downarrow 2 \!\!\! \uparrow \rangle)(|4 \!\!\! 
\uparrow 3 \!\!\! \downarrow \rangle - |4 \!\!\! \downarrow 3 \!\!\! \uparrow 
\rangle)$ and $|v \rangle = \alpha^2 (|4 \!\!\! \uparrow 1 \!\!\! \downarrow 
\rangle - |4 \!\!\! \downarrow 1 \!\!\! \uparrow \rangle)(|3 \!\!\! \uparrow 2 
\!\!\! \downarrow \rangle - |3 \!\!\! \downarrow 2 \!\!\! \uparrow \rangle)$ 
that $o = \langle h | v \rangle = 2 \alpha^4$ = 1/2. 

The single most important generic feature of raising the number of dimer 
colors, noted already in the context of $1/N$ expansions,\cite{rmr} is 
expected to be that $\alpha$ is reduced, perhaps by a power. However, 
this hypothesis is dependent on the model, and for a model such as the 
current one, where electron orbital color sectors change with bond 
direction, the overlap matrix is not uniform, and in some cases is 
identically zero. For the three cases of interest here, this can be 
shown very simply by repeating the exercise above for (os/st), (ss/ot), 
and (ss/$\gamma\gamma$) units, where $\gamma$ indicates the bond color.  
For each of the two spin--orbital triplet cases there are nine possible 
overlap matrix elements to consider, not all of which will be illustrated 
explicitly. Taking a plaquette of parallel $a$ and $c$ sides, one has 
\begin{eqnarray}
|h_{++} {\rm (os/st)} \rangle & = & \alpha^2 (|1 r \!\! \uparrow 2 r \!\! 
\uparrow \rangle - |1 g \!\! \uparrow 2 g \!\! \uparrow \rangle) \nonumber \\ 
& & \;\;\;\; \times (|4 r \!\! \uparrow 3 r \!\! \uparrow \rangle - |4 g \!\! 
\uparrow 3 g \!\! \uparrow \rangle), \nonumber \\
|v_{++} {\rm (os/st)} \rangle & = & \alpha^2 (|4 g \!\! \uparrow 1 g \!\! 
\uparrow \rangle - |4 b \!\! \uparrow 1 b \!\! \uparrow \rangle) \nonumber \\ 
& & \;\;\;\; \times (|3 g \!\! \uparrow 2 g \!\! \uparrow \rangle - |3 b \!\! 
\uparrow 2 b \!\! \uparrow \rangle), \nonumber \\
|h_{00} {\rm (os/st)} \rangle & = & \alpha^4 (|1 r \!\! \uparrow 2 r \!\! 
\downarrow \rangle + |1 r \!\! \downarrow 2 r \!\! \uparrow \rangle 
\nonumber \\ & & \;\;\;\;\;\; - |1 g \!\! \uparrow 2 g \!\! \downarrow 
\rangle - |1 g \!\! \downarrow 2 g \!\! \uparrow \rangle) \nonumber \\ & & 
\;\;\;\; \times (|4 r \!\! \uparrow 3 r \!\! \downarrow \rangle + |4 r \!\! 
\downarrow 3 r \!\! \uparrow \rangle \nonumber \\ & & \;\;\;\;\;\;
 - |4 g \!\! \uparrow 3 g \!\! \downarrow \rangle - |4 g \!\! \downarrow 
3 g \!\! \uparrow), \nonumber \\
|v_{00} {\rm (os/st)} \rangle & = & \alpha^4 (|4 g \!\! \uparrow 1 g \!\! 
\downarrow \rangle + |4 g \!\! \downarrow 1 g \!\! \uparrow \rangle 
\nonumber \\ & & \;\;\;\;\;\; - |4 b \!\! \uparrow 1 b \!\! \downarrow 
\rangle - |4 b \!\! \downarrow 1 b \!\! \uparrow \rangle) \nonumber \\ & & 
\;\;\;\; \times (|3 g \!\! \uparrow 2 g \!\! \downarrow \rangle + |3 g \!\! 
\downarrow 2 g \!\! \uparrow \rangle \nonumber \\ & & \;\;\;\;\;\; 
 - |3 b \!\! \uparrow 2 b \!\! \downarrow \rangle - |3 b \!\! \downarrow 
2 b \!\! \uparrow \rangle), \nonumber \\
|h_{++} {\rm (ss/ot)} \rangle & = & \alpha^2 (|1 r \!\! \uparrow 2 g \!\! 
\downarrow \rangle - |1 r \!\! \downarrow 2 g \!\! \uparrow \rangle) \nonumber 
\\ & & \;\;\;\; \times (|4 r \!\! \uparrow 3 g \!\! \downarrow \rangle
 - |4 g \!\! \downarrow 3 g \!\! \uparrow \rangle), \nonumber \\
|v_{++} {\rm (ss/ot)} \rangle & = & \alpha^2 (|4 g \!\! \uparrow 1 b \!\! 
\downarrow \rangle - |4 g \!\! \downarrow 1 b \!\! \uparrow \rangle) \nonumber 
\\ & & \;\;\;\; \times (|3 g \!\! \uparrow 2 b \!\! \downarrow \rangle
 - |3 g \!\! \downarrow 2 b \!\! \uparrow \rangle), \nonumber \\
|h_{00} {\rm (ss/ot)} \rangle & = & \alpha^4 (|1 r \!\! \uparrow 2 r \!\! 
\downarrow \rangle - |1 r \!\! \downarrow 2 r \!\! \uparrow \rangle 
\nonumber \\ & & \;\;\;\;\;\; + |1 g \!\! \uparrow 2 g \!\! \downarrow 
\rangle - |1 g \!\! \downarrow 2 g \!\! \uparrow \rangle) \nonumber \\ & & 
\;\;\;\; \times (|4 r \!\! \uparrow 3 r \!\! \downarrow \rangle - |4 r \!\! 
\downarrow 3 r \!\! \uparrow \rangle \nonumber \\ & & \;\;\;\;\;\; 
 + |4 g \!\! \uparrow 3 g \!\! \downarrow \rangle - |4 g \!\! \downarrow 
3 g \!\! \uparrow \rangle), \nonumber \\
|v_{00} {\rm (ss/ot)} \rangle & = & \alpha^4 (|4 g \!\! \uparrow 1 g \!\! 
\downarrow \rangle - |4 g \!\! \downarrow 1 g \!\! \uparrow \rangle 
\nonumber \\ & & \;\;\;\;\;\; + |4 b \!\! \uparrow 1 b \!\! \downarrow 
\rangle - |4 b \!\! \downarrow 1 b \!\! \uparrow \rangle) \nonumber \\ & & 
\;\;\;\; \times (|3 g \!\! \uparrow 2 g \!\! \downarrow \rangle - |3 g \!\! 
\downarrow 2 g \!\! \uparrow \rangle \nonumber \\ & & \;\;\;\;\;\; 
 + |3 b \!\! \uparrow 2 b \!\! \downarrow \rangle - |3 b \!\! \downarrow 
2 b \!\! \uparrow \rangle), \nonumber \\
|h {\rm (ss/cc)} \rangle & = & \alpha^2 (|1 b \!\! \uparrow 2 b \!\! 
\downarrow \rangle - |1 b \!\! \downarrow 2 b \!\! \uparrow) \nonumber 
\\ & & \;\;\;\; \times (|4 b \!\! \uparrow 3 b \!\! \downarrow \rangle
 - |4 b \!\! \downarrow 3 b \!\! \uparrow \rangle), \nonumber \\
|v {\rm (ss/aa)} \rangle & = & \alpha^2 (|4 r \!\! \uparrow 1 r \!\! 
\downarrow \rangle - |4 r \!\! \downarrow 1 r \!\! \uparrow) \nonumber 
\\ & & \;\;\;\; \times (|3 r \!\! \uparrow 2 r \!\! \downarrow \rangle
 - |3 r \!\! \downarrow 2 r \!\! \uparrow \rangle), \nonumber 
\end{eqnarray} 
and similarly, whence
\begin{eqnarray}
o_{++,++}{\rm (os/st)} & = & o_{--,--}{\rm (os/st)} \;\; = \;\; \alpha^4
 = 1/4, \nonumber \\
o_{+0,+0}{\rm (os/st)} & = & o_{+0,0+}{\rm (os/st)} \;\; = \;\; \alpha^6
 = 1/8, \nonumber \\
o_{0+,0+}{\rm (os/st)} & = & o_{0+,+0}{\rm (os/st)} \;\; = \;\; \alpha^6
 = 1/8, \nonumber \\
o_{00,00}{\rm (os/st)} & = & 2 \alpha^8 \;\; = \;\; 1/8, \nonumber \\
o_{0-,0-}{\rm (os/st)} & = & o_{0-,-0}{\rm (os/st)} \;\; = \;\; \alpha^6
 = 1/8, \nonumber \\
o_{-0,-0}{\rm (os/st)} & = & o_{-0,0-}{\rm (os/st)} \;\; = \;\; \alpha^6
 = 1/8, \nonumber \\
o_{\mu\nu,\rho\sigma}{\rm (os/st)} & = & 0 \;\;\;\; {\rm otherwise}, 
\nonumber \\ o_{00,00}{\rm (ss/ot)} & = & 2 \alpha^8 \;\; = \;\; 1/8, 
\nonumber \\ o_{\mu\nu,\rho\sigma}{\rm (ss/ot)} & = & 0 \;\;\;\; 
{\rm otherwise}, \nonumber \\ o{\rm (ss/\gamma\gamma)} & = & 0. 
\end{eqnarray} 
For (os/st) dimers there are finite overlap matrix elements within each $S_z$ 
sector, all (but one) of considerably higher order than the conventional 
spin--singlet problem. However, the 9$\times$9 overlap matrix is rather 
sparse. For (ss/ot) dimers, the orbital color combinations forbid any 
overlap matrix elements other than for pairs of $T_z = 0$ dimers. For 
bond--colored, spin--singlet dimers there are no overlap matrix elements 
at all. Thus the highly non--uniform nature of the matrix elements in the 
superexchange model and the fact that they vanish in the direct--exchange
model preclude any sort of meaningful expansion in the inverse of the 
overlap matrix.\cite{rrk} As noted in Sec.~I, the matrix elements $t$ and 
$v$ of the relevent QDMs will instead be deduced directly at second--order 
in perturbation theory using the electronic Hamilitonian.

\begin{figure}[t]
\includegraphics[width=4.1cm]{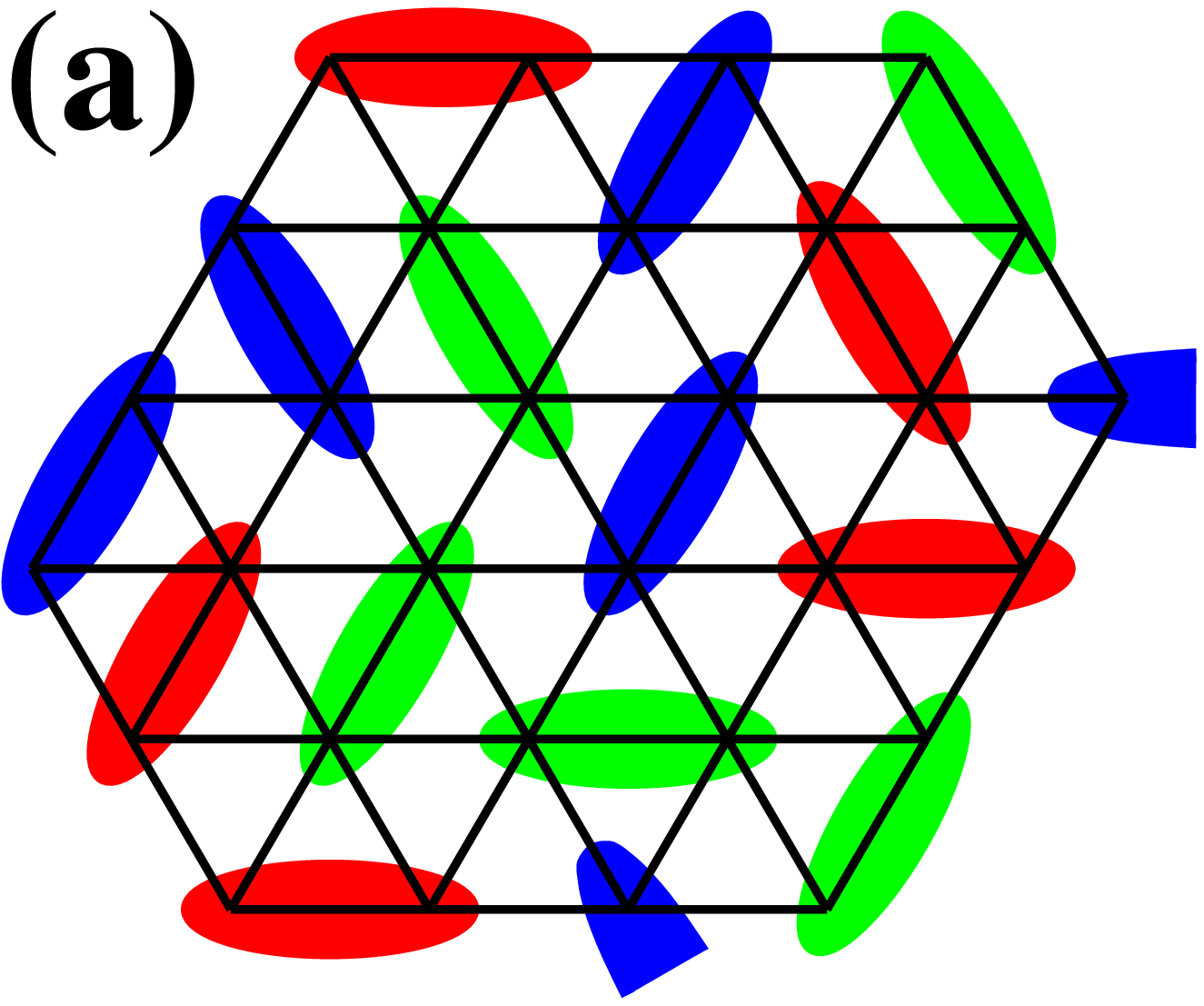}\hspace{0.2cm}\includegraphics[width=4.1cm]{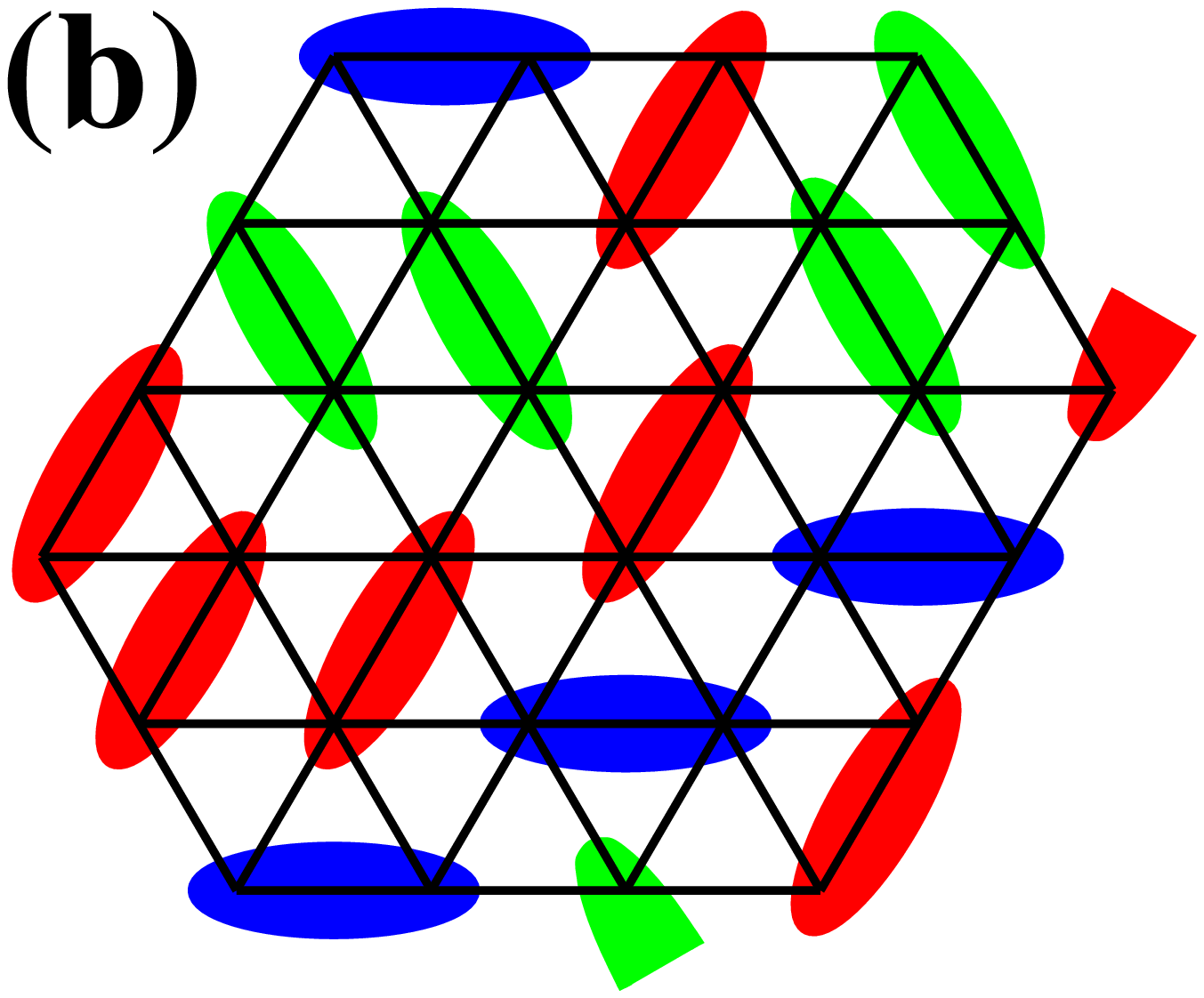}
\caption{(color online) QDM representations. (a) Three--color QDM as obtained 
in the superexchange limit for both (ss/ot) and (os/st) dimers; dimers of any 
color may occupy any bond, but not all processes ($t$ or $v$) may be allowed 
on plaquettes containing two dimers. Dimer colors do not correspond to the 
colors in Fig.~1. (b) QDM obtained in the direct--exchange limit for 
(ss/$\gamma\gamma$) dimers, whose colors do correspond to those in Fig.~1. 
Bonds of the three different orientations have only one color each, and 
thus the color is no longer a degree of freedom. This defines an effective 
one--color QDM.} 
\label{qdmf2}
\end{figure}

\subsection{Superexchange model}

In the limit $\eta = 0$, the bond Hamiltonian of Eq.~(\ref{ehs}) takes the 
more transparent form 
\begin{eqnarray}
\label{ehs0}
{\cal H}_{s0} & = &  J_s \!\!\! \sum_{\langle ij \rangle \parallel \gamma}
\!\! 2 \Big( \vec{S}_i \! \cdot \! \vec{S}_j \! + \! \frac{1}{4} \Big) 
\Big( \vec{T}_{i\gamma} \! \cdot \! \vec{T}_{j\gamma} \! + \! \frac{1}{4} 
n_i^{\gamma} n_j^{\gamma} \Big) \nonumber \\ & & \;\;\;\;\;\;\;\;\;\;\;
 + \frac12 (n_{i\gamma} + n_{j\gamma}) - \frac12. 
\end{eqnarray}
The six--fold single--bond degeneracy of (ss/ot) and (os/st) dimers is 
lifted in favor of the former by quantum fluctuations on the off--dimer 
bonds at $\eta = 0$,\cite{rno} leaving a threefold degeneracy corresponding 
to the orbital triplet states of the two electrons, $T_z = 1$, 0, and $-1$.
Thus one expects a QDM with three different flavors of dimer, a situation 
depicted in Fig.~2(a). Note that these ``colors'' are not the same as the 
spectral colors in Fig.~1, which are associated directly with the symmetry 
of the $t_{2g}$ levels, but correspond to three possible orbital states on 
each bond of a given direction [Eqs.~(\ref{eotp})--(\ref{eotm})]. The three 
analogous states for a different lattice direction are composed of a different 
$t_{2g}$--color pair. 

However, for physical values of $\eta$ around 0.1, the (os/st) dimers are 
preferred as the direct lifting of degeneracy in the one--bond Hamiltonian 
outweighs the kinetic effects. This regime is delimited by a lower bound, 
which for the four--site cluster is $\eta_l = 0.03$, and an upper bound of 
$\eta_u \sim 0.15$ where the ground state becomes ferromagnetic by a lifting 
of the spin--triplet degeneracy.\cite{rno} Within this range, one has a 
three--color QDM similar to the (ss/ot) case [Fig.~2(a)], but one in which 
the colors correspond to the spin triplet states of the two electrons, $S_z 
= 1$, 0, and $-1$. The triplet states in the spin sector are the same for 
every bond direction, and this fact is responsible for the differences from 
the (ss/ot) situation observed already in the overlap matrix elements. The 
QDM matrix elements can also be expected to differ in several respects. 

Summarizing the situation to this point, in the superexchange limit of the 
model one has a number of energetic criteria which can be compared to the 
RVB criteria of Ref.~[\onlinecite{rmvrbfp}]. The former are\cite{rno} 
({\it i}) a very strong tendency to dimer formation, ({\it ii}) a large 
semi--classical degeneracy of basis states, namely the set of coverings 
formed from these dimers, and ({\it iii}) that resonance processes on 
four--site plaquette units provide a significant energetic contribution. 
From ({\it iii}), a minimal triangular--lattice QDM should in principle 
contain, within the dimer--flipping term $t$ and the static pair term $v$, 
the leading corrections to the static VB energy (determined by $J_s$). The 
question is then reduced to whether the regime $t/v \gtrsim 1$ of RVB 
physics\cite{rms} can be attained.

\subsection{Direct--exchange model}

Here only the $\eta = 0$ limit of the model, 
\begin{eqnarray}
\label{ehd0}
{\cal H}_{d0} & = & J_d \!\! \sum_{\langle ij \rangle \parallel \gamma}
\! \Big( \vec{S}_i \! \cdot \! \vec{S}_j - \frac{1}{4} \Big) n_{i\gamma} 
n_{j\gamma} \\ & & \;\;\;\;\;\;\;\;\;\;\;\; - \frac{1}{4} \Big( n_{i\gamma} 
(1 - n_{j\gamma}) + (1 - n_{i\gamma}) n_{j\gamma} \Big), \nonumber 
\end{eqnarray}
will be considered. As noted above, the spin--singlet state optimizing 
the bond energy can be formed only from two electrons of the bond color. 
Thus the three apparent $t_{2g}$ colors of the dimers are locked rigidly 
to the bond direction, with no finite matrix elements for directional 
fluctuations, a situation depicted in Fig.~2(b). The effective QDM will 
then be equivalent to a one--color model. 

In addition to the studies of Ref.~[\onlinecite{rno}], this model has also 
been considered in some detail by Jackeli and Ivanov.\cite{rji} These authors 
demonstrated that the extensively degenerate manifold of (bond--colored) dimer 
coverings forms a set of exact eigenstates of the second--order Hamiltonian. 
In deducing the leading perturbations which would select a type of order, 
they allowed finite values of $J_H$, which lead at order $\eta^3$ to a 
ferromagnetic interaction on interdimer bonds with one (but not two) dimers 
of the bond color. As a result, particular static dimer configurations are 
selected (the type of state is unique, albeit with a translational and 
rotational degeneracy of 60), but there remains no contribution from the 
positional resonance of dimers. The aim here is to investigate the 
consequences of the lowest--order dimer resonance processes of a type 
which would give valid contributions to a QDM. 

\begin{figure}[t]
\includegraphics[width=6.5cm]{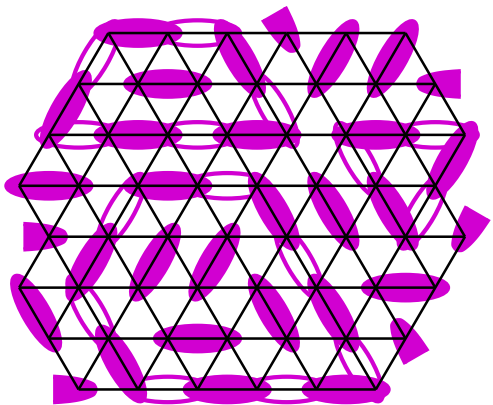}
\caption{(color online) Representation of allowed dimer fluctuation 
processes in the direct--exchange limit, shown for a system of only one 
dimer color. These processes define loops whose sides have even numbers 
of bonds, and which (to maintain complete dimer covering of the lattice) 
must enclose an even number of sites. Illustrated are the lowest three 
loops, which have lengths of 6, 10, and 14 bonds. }
\label{qdmf3}
\end{figure}

In this limit of the model, electrons in dimer singlets may hop only in the 
direction corresponding to their bond color. Thus virtual processes are 
highly restricted and the true dimensionality of the system is far lower 
than the connectivity of the triangular lattice would suggest.\cite{rno} 
Because a four--site plaquette contains no two bonds of the same 
direction, not only is the overlap matrix element zero (above) but the 
$t$ and $v$ terms of the conventional, triangular--lattice QDM are also 
identically zero. A QDM based on higher--order terms will therefore be 
required. The only kinetic contributions on the triangular lattice are 
given by paths of even side--lengths, and which enclose even numbers 
of sites. Some valid loops are shown in Fig.~3. Here the contributions 
$t'$ and $v'$ from fluctuation processes on six--bond triangles will be 
computed, while the leading perturbations from higher allowed loops (the 
10--bond trapezium and 14--bond irregular pentagon) can be argued to be 
small. The ground state of the resulting $t'$--$v'$ QDM should contain 
the leading effects of dimer resonance processes on the highly degenerate 
manifold of static VB coverings.\cite{rno} One may then also consider 
whether the energy of this state is more favorable than the spontaneous 
one--dimensionalization of the system into effectively decoupled 
Heisenberg chains of a single orbital color. This outcome, the 
lowest--energy possibility found in Ref.~[\onlinecite{rno}], is 
generally held to be rather unlikely, although from the 1D nature 
of the electronic processes in this limit of the model it is not 
entirely implausible. 

Commenting briefly on further models which may arise within the $t_{2g}$ 
system on the triangular lattice, clearly there are many of these even at 
unit filling per site. The entire range $0 < J'/J < \infty$ is of obvious 
physical importance, and in principle presents a model with four types of 
dimer. However, the lack of symmetry between the two limits implies 
automatic selection effects which will favor a smaller number of these 
dimers (those maximizing virtual kinetic proceses). The results of 
Ref.~[\onlinecite{rno}] suggest a first--order transition from the 
superexchange regime (QDMs with up to three colors) to the direct--exchange 
regime (a one--color QDM) at an intermediate value of $J'/J$ but with no 
intermediate phase. A second possibility is the regime of very high $\eta$, 
where for all values of $J'/J$ the preference for dimers will melt, and one 
obtains very highly degenerate sets of color configurations among which 
the (now subdominant) Heisenberg interactions are responsible for 
selection effects. 

To conclude this section, specific colored QDMs can be obtained for the 
three cases of interest and, within the approximations inherent in the 
derivation process (by which is meant the breaking of translational 
symmetry and the discarding of very many degrees of freedom), all are 
expected to yield additional information concerning the physics of the 
ground state of the starting electronic Hamiltonian. Before proceeding 
to a detailed analysis of these models (Sec.~IV), it is necessary to consider 
the physical possibilities contained within QDMs possessing an additional 
color degree of freedom. 

\begin{figure}[t]
\includegraphics[width=4.5cm]{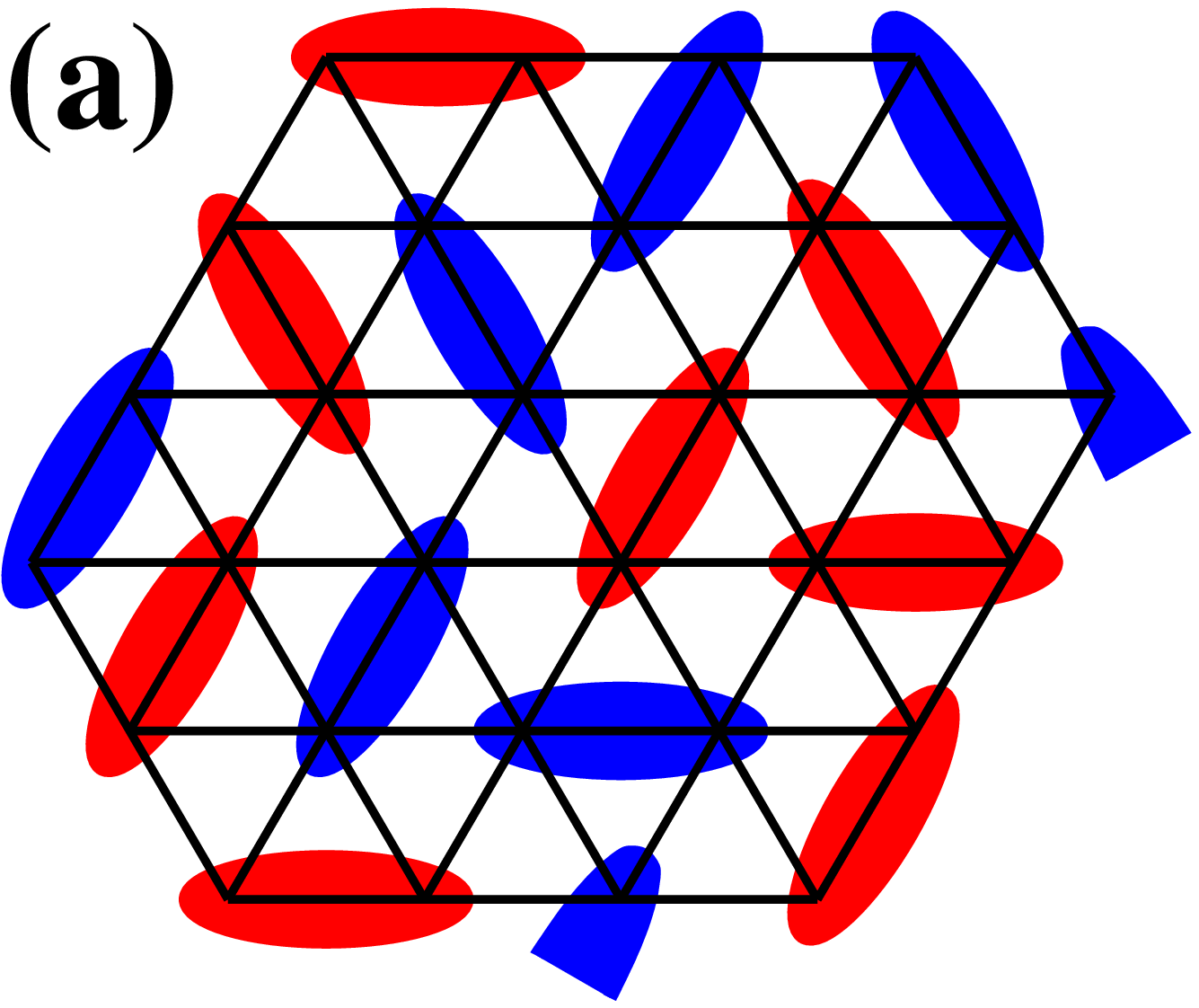} \\
\phantom{.}
\includegraphics[width=6.5cm]{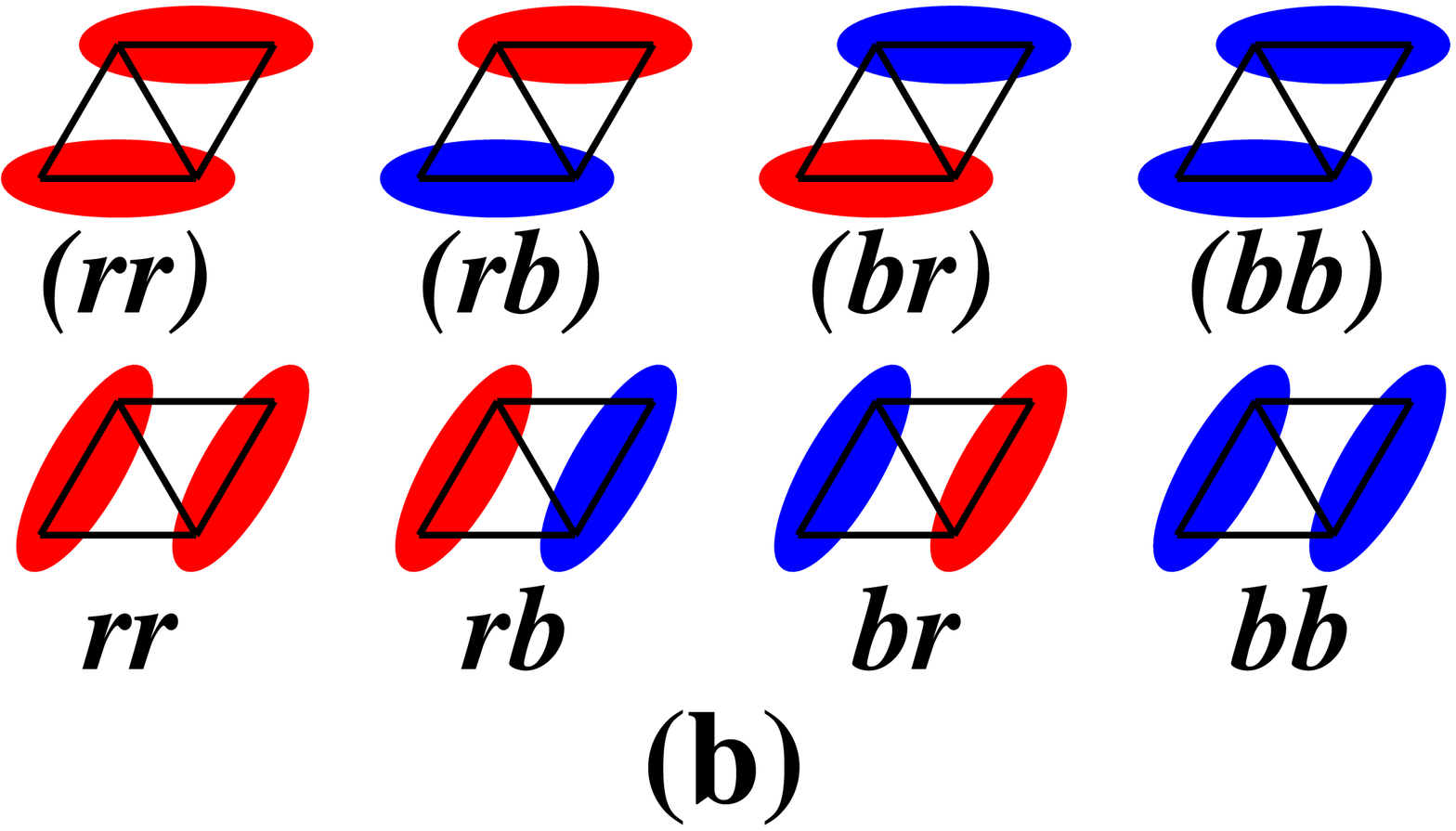}
\caption{(color online) (a) Dimer covering of the two--color QDM on the 
triangular lattice. (b) Representation of the eight possible configurations 
of two colored dimers on a flippable $(ac)$ plaquette.}
\label{qdmf4}
\end{figure}

\section{General colored QDM} 

Consider a QDM on the triangular lattice with two different types of dimer, 
which will be labeled by the colors red and blue. An arbitrarily chosen 
dimer covering is shown in Fig.~4(a). To systematize the description of this 
system, one begins by noting that there are eight possible states on each 
rhombic plaquette: taking an $(ac)$ plaquette for illustration [Fig.~4(b)], 
let these be denoted by $i = rr$, $rb$, $br$, and $bb$ for pairs of 
$a$--axis dimers and $(rr)$, $(rb)$, $(br)$, and $(bb)$ for pairs of 
$c$--axis dimers. The QDM Hamiltonian has the form 
\begin{equation}
H = \sum_{\gamma = 1}^3 \sum_{i = 1}^{2N} H_p
\label{ehqdmp}
\end{equation}
where there are $2N$ plaquettes of any one orientation for a system of $2N$ 
sites ($N$ dimers), and $\gamma$ labels the three plaquette orientations on 
the triangular lattice. The plaquette Hamiltonian is given by 

\begin{figure}[h]
\newcommand{\lb}[1]{\raisebox{-0.8ex}[0.8ex]{#1}}
\begin{center}
$H_p = - t_{rr,rr} \big( | \!$ \lb{\resizebox{0.03\textwidth}{!}{
\includegraphics[height=5cm]{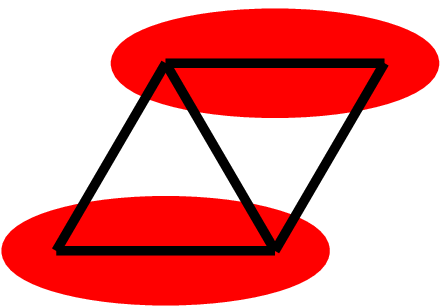}}} $\! \rangle \langle \!$
\lb{\resizebox{0.03\textwidth}{!}{
\includegraphics[height=5cm]{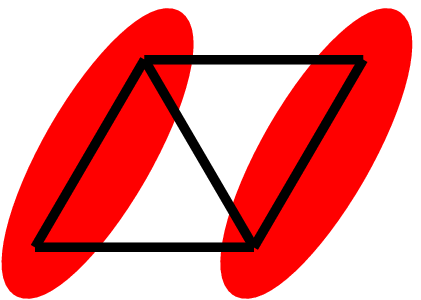}}} $\! | + | \!$
\lb{\resizebox{0.03\textwidth}{!}{
\includegraphics[height=5cm]{qdmfe2.eps}}} $\! \rangle \langle \!$
\lb{\resizebox{0.03\textwidth}{!}{
\includegraphics[height=5cm]{qdmfe4.eps}}} $\! | \big) \;\;\;\;\;\;$ \\[10pt]
$\;\;\; - t_{rr,rb} \big( | \!$ \lb{\resizebox{0.03\textwidth}{!}{
\includegraphics[height=5cm]{qdmfe4.eps}}} $\! \rangle \langle \!$
\lb{\resizebox{0.03\textwidth}{!}{
\includegraphics[height=5cm]{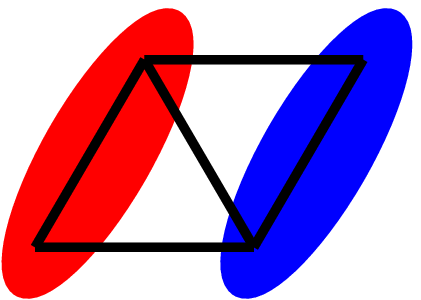}}} $\! | + | \!$
\lb{\resizebox{0.03\textwidth}{!}{
\includegraphics[height=5cm]{qdmfe7.eps}}} $\! \rangle \langle \!$
\lb{\resizebox{0.03\textwidth}{!}{
\includegraphics[height=5cm]{qdmfe4.eps}}} $\! | \big)$ \\[10pt]
$ \;\;\;\;\;\;\;\;\;\;\;\;
 - t_{rr,rb} \big( | \!$ \lb{\resizebox{0.03\textwidth}{!}{
\includegraphics[height=5cm]{qdmfe2.eps}}} $\! \rangle \langle \!$
\lb{\resizebox{0.03\textwidth}{!}{
\includegraphics[height=5cm]{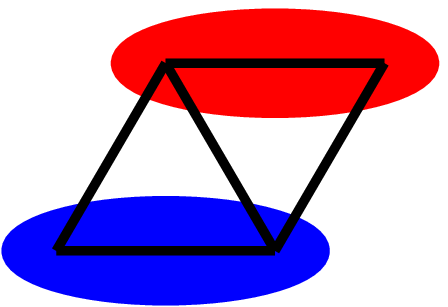}}} $\! | + | \!$
\lb{\resizebox{0.03\textwidth}{!}{
\includegraphics[height=5cm]{qdmfe8.eps}}} $\! \rangle \langle \!$
\lb{\resizebox{0.03\textwidth}{!}{
\includegraphics[height=5cm]{qdmfe2.eps}}} $\! | \big) + \dots$ \\[10pt]
$ \;\;\; - t_{rb,br} \big( | \!$ \lb{\resizebox{0.03\textwidth}{!}{
\includegraphics[height=5cm]{qdmfe8.eps}}} $\! \rangle \langle \!$
\lb{\resizebox{0.03\textwidth}{!}{
\includegraphics[height=5cm]{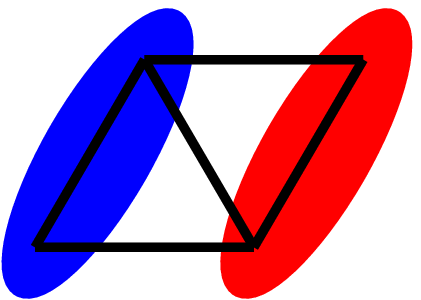}}} $\! | + | \!$
\lb{\resizebox{0.03\textwidth}{!}{
\includegraphics[height=5cm]{qdmfe6.eps}}} $\! \rangle \langle \!$
\lb{\resizebox{0.03\textwidth}{!}{
\includegraphics[height=5cm]{qdmfe8.eps}}} $\! | \big)$ \\[10pt]
$ \;\;\;\;\;\;\;\;\;\;\;\;
 - t_{rb,br} \big( | \!$ \lb{\resizebox{0.03\textwidth}{!}{
\includegraphics[height=5cm]{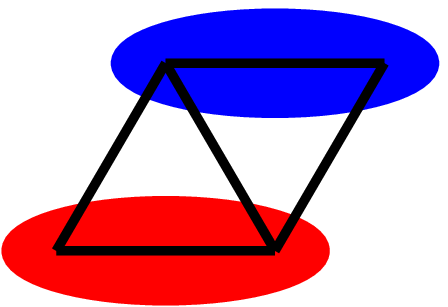}}} $\! \rangle \langle \!$
\lb{\resizebox{0.03\textwidth}{!}{
\includegraphics[height=5cm]{qdmfe7.eps}}} $\! | + | \!$
\lb{\resizebox{0.03\textwidth}{!}{
\includegraphics[height=5cm]{qdmfe7.eps}}} $\! \rangle \langle \!$
\lb{\resizebox{0.03\textwidth}{!}{
\includegraphics[height=5cm]{qdmfe5.eps}}} $\! | \big) + \dots$ \\[10pt]
$ \;\;\;\; + v_{rr,rr} \big( | \!$ \lb{\resizebox{0.03\textwidth}{!}{
\includegraphics[height=5cm]{qdmfe4.eps}}} $\! \rangle \langle \!$
\lb{\resizebox{0.03\textwidth}{!}{
\includegraphics[height=5cm]{qdmfe4.eps}}} $\! | + | \!$
\lb{\resizebox{0.03\textwidth}{!}{
\includegraphics[height=5cm]{qdmfe2.eps}}} $\! \rangle \langle \!$
\lb{\resizebox{0.03\textwidth}{!}{
\includegraphics[height=5cm]{qdmfe2.eps}}} $\! | \big)$ \\[10pt]
$ \;\;\;\;\;\;\;\;\;\;\;\;\;
 + v_{rr,rb} \big( | \!$ \lb{\resizebox{0.03\textwidth}{!}{
\includegraphics[height=5cm]{qdmfe4.eps}}} $\! \rangle \langle \!$
\lb{\resizebox{0.03\textwidth}{!}{
\includegraphics[height=5cm]{qdmfe8.eps}}} $\! | + | \!$
\lb{\resizebox{0.03\textwidth}{!}{
\includegraphics[height=5cm]{qdmfe2.eps}}} $\! \rangle \langle \!$
\lb{\resizebox{0.03\textwidth}{!}{
\includegraphics[height=5cm]{qdmfe7.eps}}} $\! | \big) + \dots$ \\[10pt]
$ \;\;\;\;\;\;\;\;\;\;\;\;\;
 + v_{br,bb} \big( | \!$ \lb{\resizebox{0.03\textwidth}{!}{
\includegraphics[height=5cm]{qdmfe5.eps}}} $\! \rangle \langle \!$
\lb{\resizebox{0.03\textwidth}{!}{
\includegraphics[height=5cm]{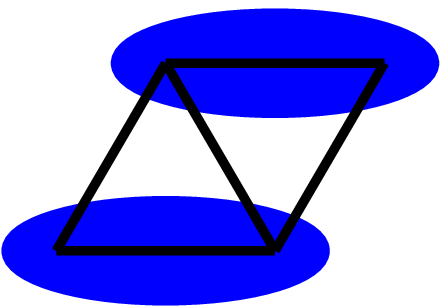}}} $\! | + | \!$
\lb{\resizebox{0.03\textwidth}{!}{
\includegraphics[height=5cm]{qdmfe6.eps}}} $\! \rangle \langle \!$
\lb{\resizebox{0.03\textwidth}{!}{
\includegraphics[height=5cm]{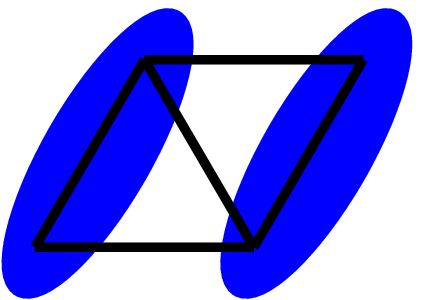}}} $\! | \big) + \dots$ 
\end{center}
\end{figure}

\noindent
and thus is represented by an 8$\times$8 matrix acting on the state vector 
$[rr,rb,br,bb,(rr),(rb),(br),(bb)]$. This matrix has a ``diagonal block'' 
$v_{ij}$ and off--diagonal blocks $t_{ij}$ and $t_{ij}^*$. The 4$\times$4 
$v_{ij}$ matrix specifies the static energies of the $rr$, $rb$, $br$, and 
$bb$ configurations, and the assumption of plaquette symmetry between $a$-- 
and $c$--axis dimer pairs, enforced here, means that this symmetric matrix 
also defines the analogous energies for $(rr)$, $(rb)$, $(br)$, and $(bb)$. 
The 4$\times$4 $t_{ij}$ matrix specifies the flipping energies of $rr$, 
$rb$, $br$, and $bb$ into $(rr)$, $(rb)$, $(br)$, and $(bb)$, and 
conversely; the symmetry $t_{i(j)} = t_{(j)i}$ (Hermitian conjugation) 
ensures that $H_{ij}$ is a symmetric matrix and the symmetry $t_{i(j)}
 = t_{(i)j}$ (one--plaquette axis permutation) ensures further that 
$t_{ij}$ is symmetric. Thus the brackets are redundant in the labels of 
$v_{ij}$ and $t_{ij}$. 

The nature of the two--color triangular--lattice QDM depends on the matrix 
elements of $v_{ij}$ and $t_{ij}$, which could in principle vary between the 
limits 
\begin{eqnarray}
v_{ij} & = & v [1,0,0,0;0,0,0,0;0,0,0,0;0,0,0,1], \nonumber \\
t_{ij} & = & t [1,0,0,0;0,0,0,0;0,0,0,0;0,0,0,1],
\label{emems}
\end{eqnarray} 
and 
\begin{eqnarray}
v_{ij} & = & v [1,1,1,1;1,1,1,1;1,1,1,1;1,1,1,1], \nonumber \\
t_{ij} & = & t [1,1,1,1;1,1,1,1;1,1,1,1;1,1,1,1].
\label{ememt}
\end{eqnarray} 
The system represented by Eqs.~(\ref{emems}) is one where potential and 
kinetic energy are gained only on plaquettes containing two dimers of the 
same color, and hence this is a dilute version of the one--color QDM. While 
some additional symmetries may arise (below) from the conservation of each 
color individually, the energetics and dynamics of the system appear to be 
somewhat trivial as a result of the dilution. Equations (\ref{ememt}) 
represent a ``color--blind'' system where the standard QDM processes occur 
with the same amplitudes irrespective of dimer color, and is therefore 
equivalent to a one--color QDM but with a redundant degree of freedom. 
While it is the signs of the elements of $v_{ij}$ and $t_{ij}$ which are 
most important in determining the ground state of the model, other than 
$v_{ii}$ there is in general no need for these to be real, and one could 
consider matrix elements which have different signs or phase factors 
occurring in different patterns.

\subsection{Topological structure}

One of the most important properties of QDMs is the fact that they display 
a topological structure (also ``topological order''),\cite{rmr} which is 
characterized by the presence of topological invariants and associated gauge 
symmetries. For a colored QDM it is therefore natural to investigate whether 
there exists some additional, nontrivial topological structure arising from 
the presence of the color degree of freedom.

For this one must first recall the origin of the $Z_2$ gauge symmetry which 
exists in the conventional QDM. This is ultimately a consequence of the 
condition that each site must contain only one hard dimer on one of the 
bonds connected to it. If one considers the number of dimers intersecting
an arbitrary line which spans the system, it is this constraint which is 
responsible for the fact that any dimer--rearrangement process (which 
defines a loop) can change the number of dimers intersecting the line only 
in multiples of 2. This property cannot be affected by any set of color 
matrix elements $t_{ij}$ and $v_{ij}$ within the framework presented above, 
and therefore this $Z_2$ symmetry cannot be embedded within some higher 
symmetry group. Alternatively stated, while it is possible to alter dimer 
colors without altering the number which intersect the arbitrarily chosen 
line, it is not possible to alter the number intersecting the line without 
altering the numbers of dimers of each color which intersect it. 

Thus the symmetry of the colored QDM can be at best $Z_2$$\otimes$$G$, where 
$G$ is a different symmetry group. A number of possibilities could be 
considered which avoid this hard constraint, including ``soft'' dimers 
which must only have unit squared amplitude on the sum of the bonds at a 
site, multiple dimers per site (for example a representation of a $S = 1$ 
spin system), a diluted model with undimerized sites or pairs of sites, 
and models of quantum trimers or quantum quadrumers on particular lattices 
and with particular matrix elements analogous to $t_{ij}$ and $v_{ij}$. 
While some of these may indeed be constructed to deliver topological 
structures and gauge symmetries beyond $Z_2$, they will not be considered 
further in the present context. 

Returning to a consideration of the number of dimers intersecting an 
arbitrary line across the system, the quantity $N = N_r + N_b$ (the 
total numbers of red and blue dimers in the system) is always a constant, 
while $n_r + n_b$ (the numbers of red and blue dimers intersecting the 
line) may change only in units of 2. This property defines the four 
topological sectors on a torus and is reflected in the $Z_2$ symmetry 
discussed above. 

Nontrivial color sectors are characterized by $n_r - n_b$. This quantity 
can vary from $-N$ to $+N$ in steps of 2; there are $N$ sectors if $N$ is 
odd and $N + 1$ if $N$ is even. Thus a trivial color--sector structure, 
where sectors with $(N_r,N_b) = (K,L)$ are connected to sectors with 
$(K \pm 1,L \mp 1)$, is characterized by $(n_r - n_b)/2$ changing in 
steps of 1. Two nontrivial alternatives arise in the two--color QDM.
\begin{itemize}

\item[(i)] If the matrices $t_{ij}$ and $v_{ij}$ have the forms $[a,0,0,0;
0,b,c,0;0,c,b,0;0,0,0,d]$, $N_r$ and $N_b$ are conserved individually 
and $(n_r - n_b)/2$ is a constant. Thus there are $N$ or $N + 1$ topological 
color sectors which cannot be mixed. The symmetry may be labeled 
$Z_2$$\otimes$$\mathbf{Z}$. Models with an infinite set of discrete 
topological invariants labeled by the set of integers $\mathbf{Z}$ 
constitute a realization of an ensemble of string nets, the theory of 
which can be found in Ref.~[\onlinecite{rw}].

\item[(ii)] If the matrices $t_{ij}$ and $v_{ij}$ have the forms $[a,0,0,e;
0,b,c,0;0,c,b,0;e,0,0,d]$, terms interconverting between two red and two 
blue dimers are also permitted. In this case, sectors $(K,L)$ are connected 
to sectors $(K \pm 2,L \mp 2)$ and $(n_r - n_b)/2$ (mod 2) is a topological 
invariant. Sectors $(K,L)$ and $(K \pm 1,L \mp 1)$ have no overlap, and 
changing the color of a single dimer is a process which results in a 
topological defect (below), one which cannot be rectified without a 
string of local processes which extends to the boundary of the system. 
In this case one has four additional color sectors on a torus, and the 
symmetry is $Z_2$$\otimes$$Z_2$: one $Z_2$ is for the hard dimer structure 
and the other for the color structure. 

\end{itemize}
If further 0 elements in case (ii) were to be made finite, then the 
Hamiltonian matrix would mix sectors differing by $(n_r - n_b)/2 = 1$, 
and the trivial topological structure would be restored.  

Thus the presence of a dimer color in a QDM can lead to additional topological 
structure in the color sector. 

\subsection{RK points and RVB phases}

Another key property of QDMs is the fact that they display Rokhsar--Kivelson 
(RK) points. A further essential question for colored QDMs therefore concerns 
the existence of RK points in models described by the different types of 
matrix above. The RK point of the original QDM occurs at $t = v$. It has the 
property that the equal--amplitude superposition (with equal phase) of all 
possible dimer coverings in each topological sector, a ``liquid'' state with 
no local order, is the ground state. While on the square lattice there are 
gapless excitations at the RK point, on the triangular lattice this ground 
state is completely gapped and the properties of the RK point are preserved 
across a phase of finite extent. This is the RVB phase. 

The equal--amplitude superposition $|\psi \rangle = {\textstyle \frac{1}
{\sqrt{N}}} \sum_1^{N_c} |c \rangle$, where $|c \rangle$ represents the 
$N_c$ dimer coverings in the chosen topological sector, has energy $\langle 
\psi |H| \psi \rangle = (v - t) \langle n^{\rm fl} \rangle$, where $n^{\rm 
fl}$ is the number of flippable plaquettes in a covering. To show that this 
wave function is an eigenstate requires writing the QDM Hamiltonian as a 
sum of plaquette projectors, which in turn requires the condition $v = t$. 
At the RK point, the energy is zero, while on the triangular lattice it is 
negative over a range of values $1 \le t/v \lesssim 1.3$, beyond which there 
are transitions to different ground states (of ordered dimers or groups of 
plaquettes, hence breaking the translational and rotational symmetry of the 
lattice).

The proof that a more general $|\psi \rangle$ is an eigenstate of a more 
general QDM Hamiltonian sets the restrictive condition that $H$ must still 
be expressible as a sum of projectors. In a two--color model of the type 
considered in this section, where $H$ contains many terms $t_{ij}$ and 
$v_{ij}$, the RK condition is that $t_{ij} = v_{ij}$ for each matrix 
component. The wave function in this case contains all coverings, in each 
topological sector (where now the color may increase the number of sectors), 
of dimers of both colors. If this structure is not maintained, $|\psi 
\rangle$ will not be an eigenstate, but may still be the ground state, 
and will retain a gap to all excitations up to the boundary of the RVB 
phase. 

The condition that $|\psi \rangle$ be the ground state of $H$ is, up to 
a point, less strict. The energy of the two--color version of the wave 
function $|\psi \rangle$ is 
\begin{equation}
E = \sum_{ij} g_{ij} (v_{ij} - t_{ij}), 
\label{egrkc}
\end{equation}
where $g_{ij}$ is a statistical factor accounting for the probability of 
finding each specific plaquette state (here those plaquettes not only 
containing two dimers but also with a finite matrix element in $t_{ij}$ 
and/or $v_{ij}$ for processes involving these). In fact $g_{ij} = \langle 
n_{ij}^{\rm fl} \rangle$ is the average number of flippable plaquettes of 
each type in the equal--amplitude superposition. For equal dimer numbers, 
$N_r = N_b$, in a two--color model, all coefficients $g_{ij}$ are equal, 
but this is no longer the case if, for example, $N_r \ne N_b$, or if a 
more complex model defined on a more complex geometry is considered. 

The physics of the colored QDM is in this regard not qualitatively 
different from that of the conventional QDM. The equal--amplitude 
superposition does not gain a large amount of energy on each plaquette, 
but it is the ground state because it profits from quantum fluctuations 
on every single plaquette, and not merely from a symmetry--broken subset 
of these (which can be at most 1/6 of the total plaquette number). The 
freedom to have every component obey $1 \le t_{ij}/v_{ij} \lesssim 1.3$ 
clearly defines a relatively broad regime of parameter space: should the 
ratio stray outside these bounds for some components, the preference of 
the remaining components for full resonance will act to maintain the RVB 
phase. 

\begin{figure}[t]
\includegraphics[width=7.5cm]{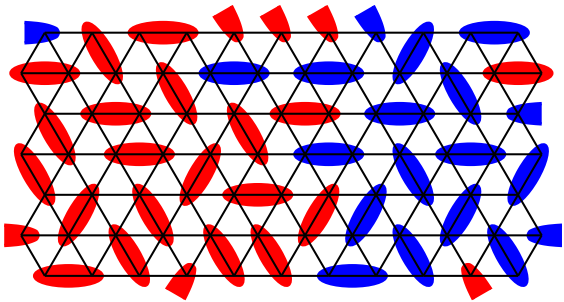}
\caption{(color online) Schematic representation of a two--color 
dimer system undergoing phase separation: if $t_{rr,rr} > v_{rr,rr}$ 
but $t_{bb,bb} < v_{bb,bb}$, a dynamically driven phase separation 
will occur (mitigated by larger values of matrix elements mixing with 
local $rb$ and $(rb)$ states). The ``staggered'' configuration of the 
predominantly blue region is static, whereas the red region undergoes 
local, symmetry--restoring resonance processes. } 
\label{qdmps}
\end{figure}

However, in this situation a multicolored QDM is susceptible to phase 
separation. As a specific but simple example, consider a model with only 
the terms $v_{rr,rr}$, $v_{bb,bb}$, $t_{rr,rr}$, and $t_{bb,bb}$ (all 
positive) large, while other matrix elements are finite [to avoid the 
dilution paralysis problem of Eq.~(\ref{emems})] but very small. The 
large elements obey $t_{rr,rr} > v_{rr,rr}$ and $t_{bb,bb} < v_{bb,bb}$, 
whence even if 
\begin{equation}
t_{rr,rr} - v_{rr,rr} > v_{bb,bb} - t_{bb,bb},
\label{egpse}
\end{equation}
so that the sum of kinetic terms exceeds that of potential terms, flippable 
blue plaquettes will entail a local energy cost. This can be avoided if blue 
dimers tend to adopt a ``staggered'' phase,\cite{rms} meaning any one of 
the set of dimer coverings ensuring that there are no flippable blue 
plaquettes. The transition to the staggered phase in the conventional 
triangular--lattice QDM is of first order, making the tendency towards phase 
separation strong. In the meantime, red plaquettes may continue to profit 
from resonance across a fraction of the system consistent with the ratio 
$N_r/N$ (Fig.~5), but the wave function is now far from an equal--amplitude 
superposition of all possibilities. At the opposite boundary of the RVB 
region, one of the colors could be expected to begin to cluster in the 
$\sqrt{12} \! \times \! \sqrt{12}$ formation.\cite{rms,rrfbim} This 
phase--separation effect becomes less pronounced as the values of the 
other matrix elements in $H_{ij}$ (for example contributions from 
flippable red--blue plaquettes) are increased. 

\subsection{Color visons and colored Majorana fermions}

The presence of topologically distinct $Z_2$ color sectors in case (ii) 
above results in color vison excitations. While local processes 
contained in the matrix elements of $t_{ij}$ and $v_{ij}$ result in 
excited states in the same topological sector, some states with very 
minimal differences can be in different topological sectors. An example 
in case (ii) is the exchange of color on a single dimer, a process not 
permitted by the local matrix elements. The (low--lying) excitation 
which results is topological in nature: to connect the initial and 
final states requires a string of local processes which spans the 
system.\cite{riif} An alternative, similar to an intermediate stage 
in the process of ``repairing'' the topological defect, is the presence 
of defect pairs, or color visons connected by a finite--length path. These 
defects are exactly analogous to the conventional $Z_2$ visons which arise 
from the dimer positions, but here they are (for a pure color vison, and 
not a vison of mixed character) present only in the color sector, {\it 
i.e.}~without altering any dimer positions. Topological excitations will 
also be present in the $\mathbf{Z}$ color sector, for which the reader is 
referred to Ref.~[\onlinecite{rw}]. 

Statements about color visons may be made rigorous by formulating a 
Majorana-fermion representation of the statistical average over the 
different dimer coverings. Following Ref.~[\onlinecite{riif}], the 
complete ensemble of dimer configurations may be represented by the 
partition function of a set of real, auxiliary fermionic variables 
$a_l^m$ corresponding to each lattice site, 
\begin{equation}
Z = \int \! \prod_{l,m} da_l^m \exp \! \big[ \!\! \sum_{ll'mm'} \!\! 
a_l^m A_{ll'}^{mm'} a_l^{m'} \big] = {\rm Pfaff} (A_{ll'}^{mm'}),
\label{emfpf}
\end{equation}
where the right--hand side is the Pfaffian of the matrix of effective 
Majorana--fermion hopping amplitudes on each bond $\langle ll' \rangle$ 
of the lattice. In the two--color QDM, two colors of Majorana fermion are 
required for each site, and this is represented in the index $m$. The 
matrix elements $A_{ll'}^{mm'}$ take the values $\pm 1$ for neighboring 
sites $l$ and $l'$, and 0 otherwise. Hence each bond is described by a 
2$\times$2 matrix 
\begin{equation}
\overleftrightarrow{A}_{ll'} = \left( \begin{array}{cc} A_{ll'}^{rr} & 
A_{ll'}^{rb} \\ A_{ll'}^{br} & A_{ll'}^{bb} \end{array} \right), 
\label{esmfm}
\end{equation}
which is net antisymmetric under exchange of $l$ and $l'$ ($A_{ll'}^{mm} 
 = - A_{l'l}^{mm}$, $A_{ll'}^{mm'} = - A_{l'l}^{m'm}$). Because all but the 
sparsest models contain plaquette processes which on any given bond can 
exchange red and blue dimer ends, and change the number of dimer ends of 
each color, all matrix elements are finite in the $(r,b)$ basis. The 
condition which ensures that all dimer coverings appear in the partition 
function with the correct relative sign is that the product $\prod_\Gamma 
\overleftrightarrow{A}_{ll'} = - \overleftrightarrow{I}$ on any loop $\Gamma$ 
consisting of an even number of bonds $\langle ll' \rangle$: the product 
of an even number of 2$\times$2 matrices around a plaquette (of 4 to $2N$
sites) should be the negative 2$\times$2 identity. The two $-1$ terms in 
$- \overleftrightarrow{I}$ correspond to the two $Z_2$ sectors in the 
two--color QDM.

The Pfaffian formulation becomes more transparent on symmetrizing the 
Majorana fermions, 
\begin{eqnarray}
a_l^s & = & {\textstyle \frac{1}{\sqrt{2}}} (a_l^r + a_l^b), \nonumber \\
a_l^a & = & {\textstyle \frac{1}{\sqrt{2}}} (a_l^r - a_l^b),
\label{emfs}
\end{eqnarray}
whence ${\tilde A}_{ll'}^{ss} = (A_{ll'}^{rr} + A_{ll'}^{rb} + A_{ll'}^{br}
 + A_{ll'}^{bb})/2$ and similarly for the other three elements of ${\tilde 
A}_{ll'}$ in the symmetrized basis. The situation is particularly simple in 
a color--symmetric problem, by which is meant one with $N_r = N_b$, 
$v_{rr,rr} = v_{bb,bb}$, and $t_{rr,rr} = t_{bb,bb}$; other situations 
are qualitatively similar but notationally more complex. In the 
color--symmetric case, the model is invariant under interchange of the 
two colors, $r \leftrightarrow b$. Hence $A_{ll'}^{rr} = A_{ll'}^{bb}
 = A_{ll'}^{mm}$ and $A_{ll'}^{rb} = A_{ll'}^{br} = A_{ll'}^{mm'}$, 
which leads to 
\begin{equation}
\left( \!\! \begin{array}{cc} {\tilde A}_{ll'}^{ss} & {\tilde A}_{ll'}^{sa} \\ 
{\tilde A}_{ll'}^{as} & {\tilde A}_{ll'}^{aa} \end{array} \!\! \right) = 
\left( \!\! \begin{array}{cc} A_{ll'}^{mm} \! + \! A_{ll'}^{mm'} & 0 \\ 0 & 
A_{ll'}^{mm} \! - \! A_{ll'}^{mm'} \end{array} \!\! \right) \! . 
\label{esmfms}
\end{equation}
Thus the bond matrix is diagonal in the ($r$+$b$, $r$$-$$b$) basis. The 
plaquette product then reduces to two products of even numbers of scalars, 
each of which has value $-1$. The upper product corresponds to the 
(hard--dimer--related) $Z_2$ symmetry of the conventional one--color 
model and the lower to the $Z_2$ symmetry of the color sector. 

\begin{figure}[t]
\includegraphics[width=7cm]{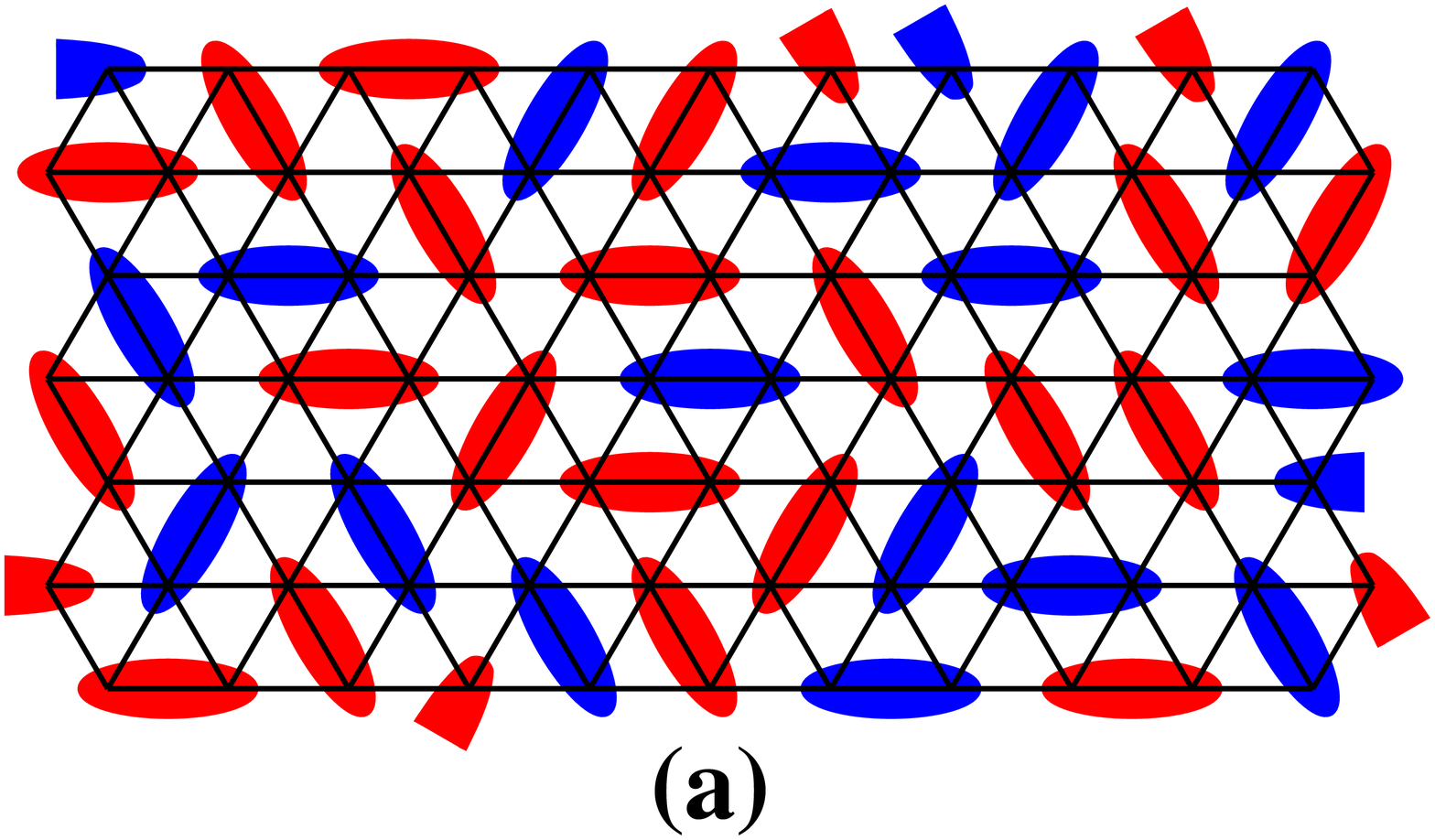}
\includegraphics[width=7cm]{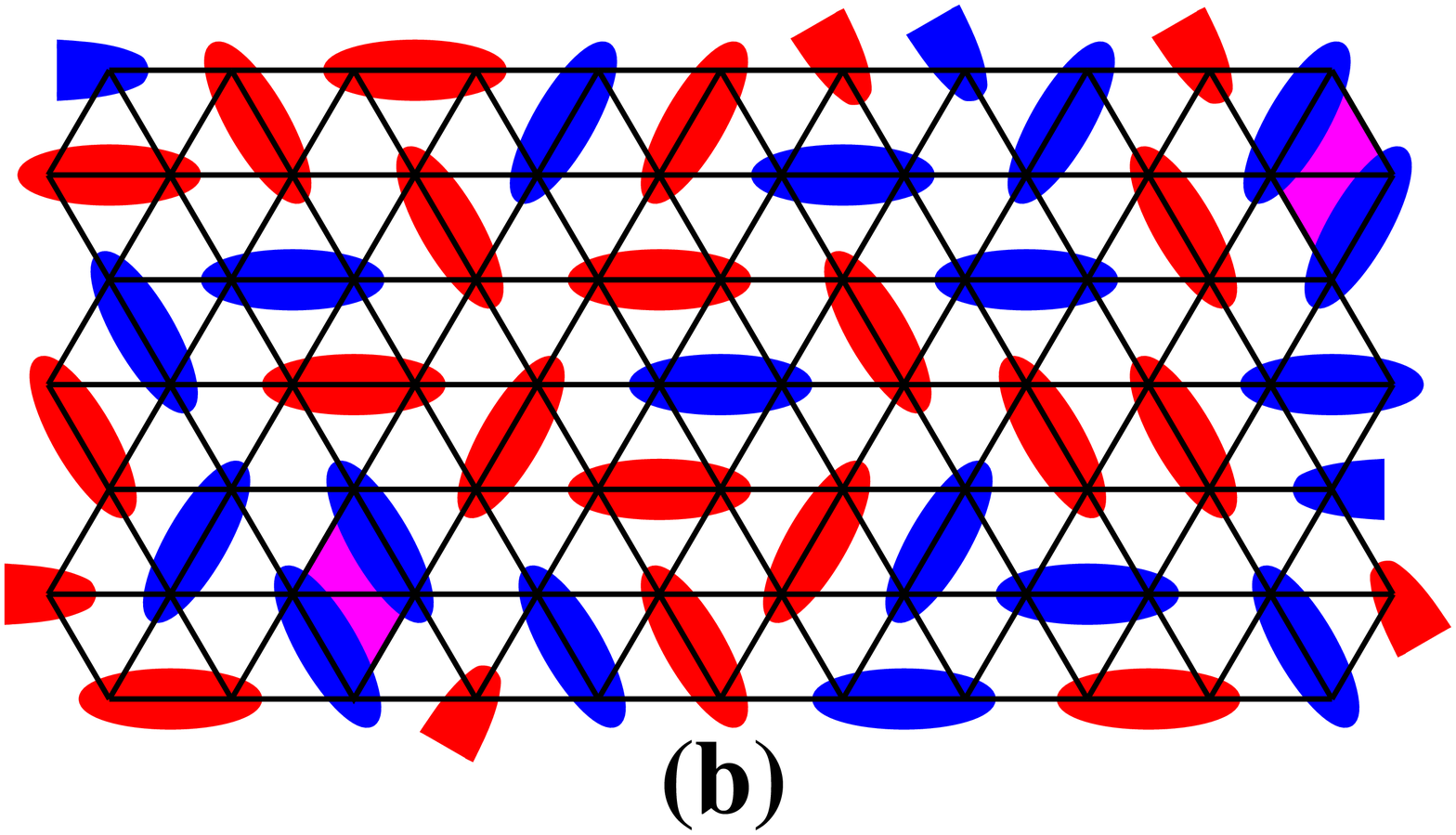}
\includegraphics[width=7cm]{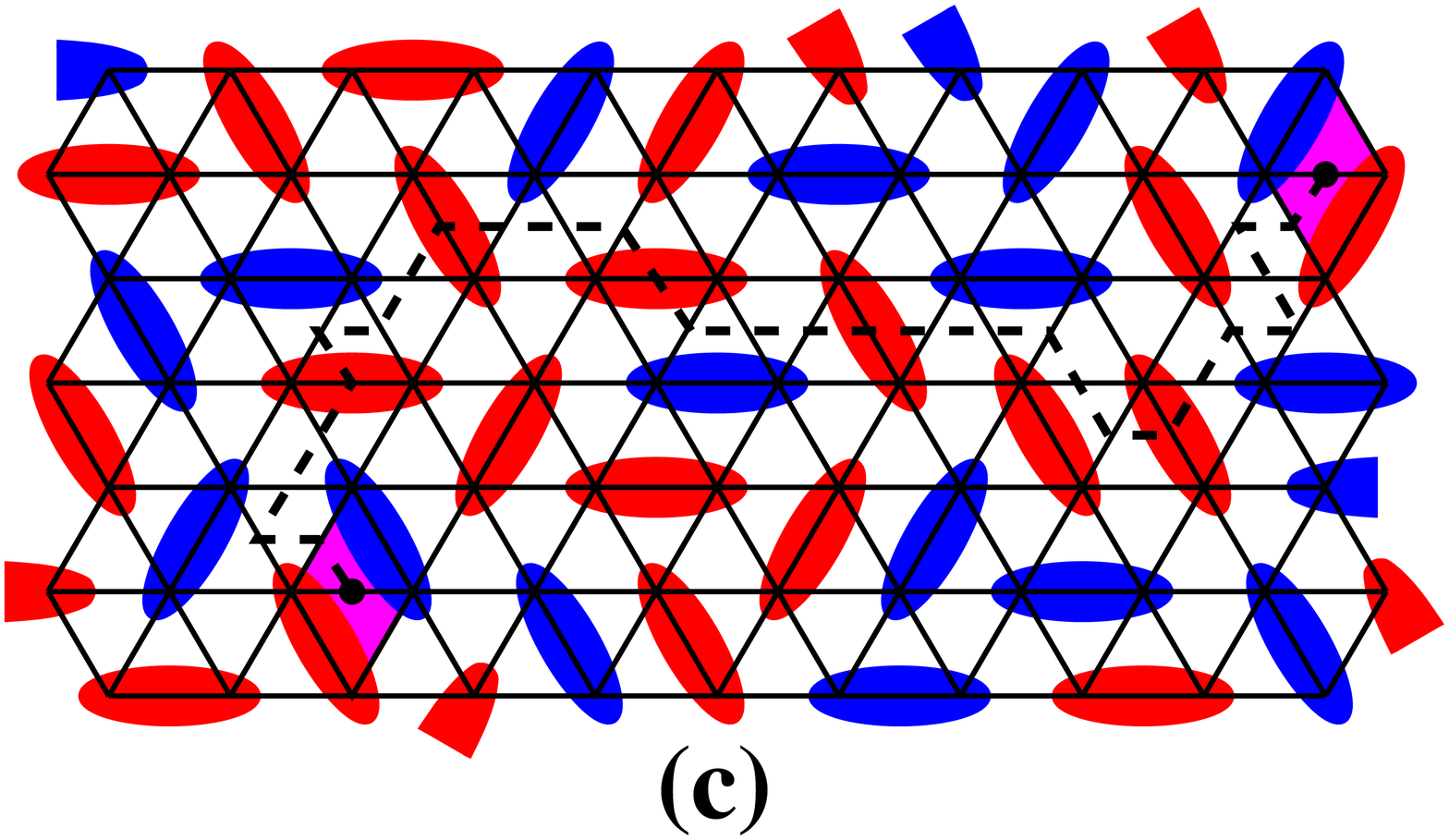}
\caption{(color online) (a) Arbitrarily chosen dimer configuration in a 
two--color QDM. (b) A color vison is introduced in the lower left corner 
by exchanging the color of a single dimer (here from red to blue) without 
altering the positions of any dimers of either color. Because this 
configuration would not be in the same topological color sector as that 
in panel (a), attempting to restore configuration (a) by applying the local 
matrix elements of case (ii) results in a path which spans the system. If, 
however, a second color vison is also present, in the upper right corner, 
then the vison pair may annihilate. For simplicity, the visons are 
represented as residing on flippable plaquettes in which the 
color--exchanged dimer is involved; dealing with color--exchanged dimers 
not initially on a flippable plaquette is a small extension, and more 
general considerations can be used to isolate visons as triangle--based 
entities.\cite{riif} (c) One possible path of local processes connecting 
the two color visons.}
\label{qdmv}
\end{figure}

Defective plaquettes, those where the product is no longer $-1$, describe 
states which do not belong in the same topological sector, and hence are 
topological defects. Defects in the upper index of the symmetrized bond 
matrix are conventional visons, while those in the lower index are color 
visons. While it is difficult (for obvious reasons) to represent a single 
vison in a finite region of a system, Fig.~6 presents a schematic 
representation of two color visons connected by a path of finite length.
The $Z_2$ symmetry of the color sector dictates that a vison is its own 
antivison. 

The discussion can be extended in a straighforward manner to three dimer 
colors. It is important to note first that three colors do not imply a 
$Z_3$ symmetry, or indeed anything similar to this. A $Z_3$ symmetry would 
require three separate topological sectors for each periodic direction on a 
cylinder or torus, and no aspect of the purely pairwise interactions in a 
QDM permits terms creating such a structure; this is not a question of the 
choice of the matrix elements in $t_{ij}$ and $v_{ij}$, but one concerning 
the architecture of the model. With green ($g$) as the third color, the 
analogs of cases (i) and (ii) above are rather the following.
\begin{itemize}

\item[(i$_3$)] For interactions conserving the number of dimers of each 
color, there are $N(N+1)/2$ (for odd $N$) or $(N+1)(N+2)/2$ (for even $N$) 
separate and unmixed color sectors which would be labeled by 
$\mathbf{Z}$$\otimes$$\mathbf{Z}$. 

\item[(ii$_3$)] For interactions conserving the number of dimer colors in 
each pair modulo 2, there is a set of ``three'' separate $Z_2$ subsectors 
corresponding to $(n_r - n_b)/2$, $(n_b - n_g)/2$ and $(n_g - n_r)/2$ (mod 
2). One of the three constraints is trivially redundant (a necessary 
consequence of the other two), and hence the model has a 
$Z_2$$\otimes$$Z_2$$\otimes$$Z_2$ topological structure with
 one hard--dimer sector and two independent $Z_2$ color sectors. 

\end{itemize}
A Pfaffian formulation of the statistical average over all dimer coverings 
would require three Majorana fermions per site. In case (ii$_3$) it would 
contain one conventional vison plus two types of $rgb$ vison.

\section{Triangular--lattice $t_{2g}$ models}

In this section, let $J = 4t_e^2/U$ denote the bare superexchange bond 
strength ($J_s$ in Sec.~II) and $J' = 4t_e'^2/U$ the bare direct-exchange 
interaction ($J_d$ in Sec.~II). Some of the results to follow lie beyond 
the mean--field Ans\"atze used in Ref.~[\onlinecite{rno}]. All derivations 
will be restricted to the lowest relevant order in the electronic 
Hamiltonian. Further matrix elements would naturally be present at higher 
orders, and although these are small, they may have qualitative effects 
such as the mixing of color sectors ({\it i.e.} the introduction of further 
off--diagonal elements in the QDM matrices).

In the results of this section, the QDM matrix elements $v$ and $v'$ are 
always negative because all of their even--order contributions are net 
energy gains. The signs of $t$ and $t'$ depend on the choice of phase for 
each singlet, and hence on an arrow--direction convention applied to the 
triangular lattice. These signs are, however, unimportant in the analysis 
of a QDM, where they are a matter of convention (easily altered by a change 
of bond phases),\cite{rrk} and thus will not be tracked explicitly. 
In fact the signs of $t$ and $t'$ here, with no further manipulation, are 
generally positive (bearing in mind the overall minus sign in the QDM 
definition), because all loops involve even numbers of plaquette edge 
bonds of the same type. The sign of the ratio $t/v$ is, however, crucial 
in determining the ground state.  

\subsection{Superexchange limit, $\eta = 0$}

Here the ground state of the system is composed of (ss/ot) color triplets. 
As noted in calculating the overlap of different dimer states in Sec.~II, 
this model has quite asymmetric matrix elements due to the breaking of 
translational symmetry involved in considering four--site plaquette units. 
This is due specifically to the direct constraint on the orbital space by 
the choice of plaquette, and hence is stronger for (ss/ot) than for (os/st) 
dimers. 

\begin{figure}[t]
\includegraphics[width=4.1cm]{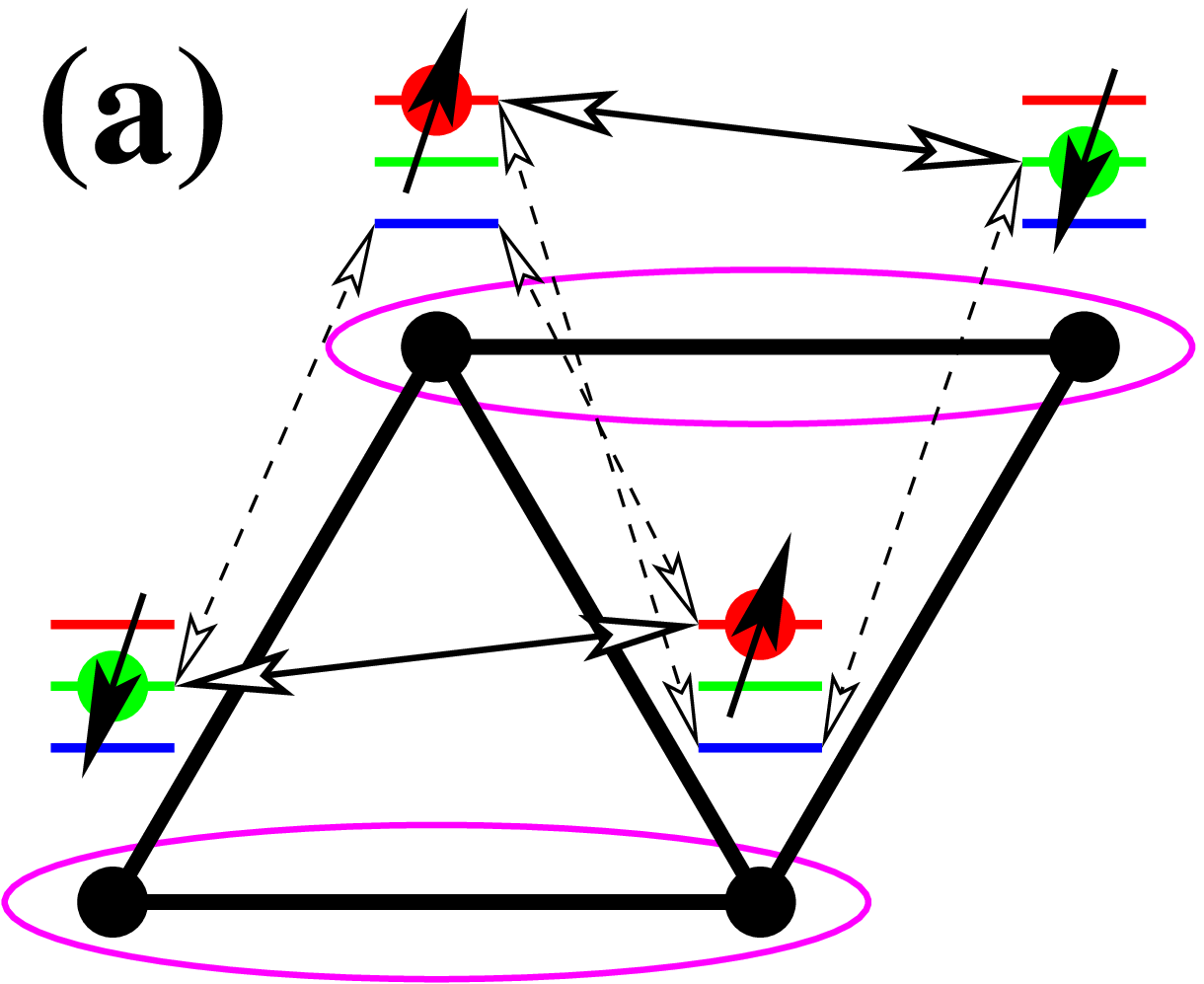}\hspace{0.2cm}\includegraphics[width=4.1cm]{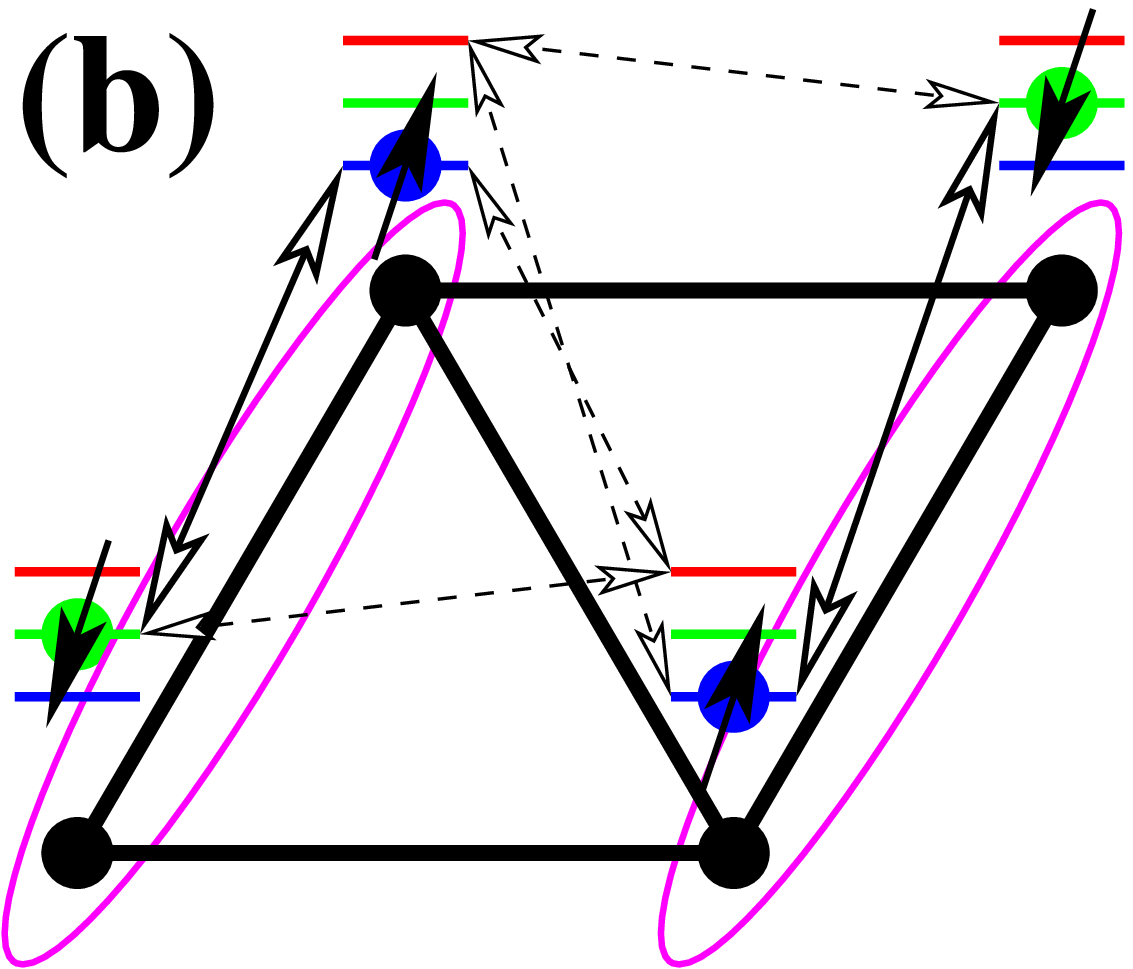}
\caption{(color online) Superexchange limit, $(ac)$ plaquette, showing 
selected (a) horizontal and (b) ``vertical'' (ss/ot) dimer pairs (magenta 
ellipses), along with a schematic representation of allowed electron hopping 
processes. The orbital triplet states are $+-$ in both panels.}
\label{qdmf11}
\end{figure}

Consider an $(ac)$ plaquette, as represented in Fig.~7(a). Let the label 
$+$ denote the state $|rg \rangle$, $-$ the state $|gr \rangle$ and 0 the 
state ${\textstyle \frac{1}{\sqrt{2}}} |rr \rangle + |gg \rangle$ [see 
Eqs.~(\ref{eotp})--(\ref{eotm})]. The allowed electron--hopping processes 
out of the red--green bond dimers are for green electrons to hop along the 
$c$--axis bonds, becoming blue as they do so, and for red electrons to hop 
along the single $b$--axis bond, also becoming blue. The change of color on 
hopping means that there is no effect of the spin state in blocking possible 
processes. 

For finite elements in the $t$ matrix, the final state must be a pair of 
spin singlets on the $c$--axis bonds with green--blue color triplet character, 
as represented in Fig.~7(b). Somewhat surprisingly, there exists precisely 
one second--order process which achieves this, namely a single exchange of 
two initially red electrons on the $b$--axis bond when both the other 
electrons on the cluster are green (this situation is represented in Fig.~7). 
There is thus no need to consider correlated hopping processes occurring on 
both $c$--axis bonds simultaneously (which are in fact available only for 
the $T_z = 0$ states of all four bonds on the plaquette), as these are of 
order $t_e^4/U^3$. Thus at second order the plaquette--flipping matrix has 
the maximally sparse and inhomogeneous form
\begin{eqnarray}
t_{+-,+-} & = & {\textstyle \frac{1}{2}} J, \nonumber \\
t_{\mu\nu,\rho\sigma} & = & 0 \;\; {\rm otherwise}. 
\label{etmess}
\end{eqnarray}
This rather extreme form is due not only to the strict constraints set by 
the available color combinations, but also to the strong breaking of 
translational symmetry encoded in the choice of rhombic plaquette. 

Under these circumstances, it is necessary to consider the second--order 
contributions to the QDM $v$ term not contained in the dimer energy itself. 
From fluctuation processes on the three interdimer bonds one obtains
\begin{eqnarray}
v_{++,++} & = & - {\textstyle \frac{3}{4}} J, \nonumber \\
v_{+0,+0} & = & - {\textstyle \frac{3}{4}} J, \nonumber \\
v_{0+,0+} & = & - {\textstyle \frac{5}{8}} J, \nonumber \\
v_{+-,+-} & = & - J, \nonumber \\
v_{00,00} & = & - {\textstyle \frac{3}{4}} J, \\
v_{-+,-+} & = & - {\textstyle \frac{1}{2}} J, \nonumber \\
v_{0-,0-} & = & - {\textstyle \frac{3}{4}} J, \nonumber \\
v_{-0,-0} & = & - {\textstyle \frac{5}{8}} J, \nonumber \\
v_{--,--} & = & - {\textstyle \frac{3}{4}} J, \nonumber \\
v_{\mu\nu,\rho\sigma} & = & 0 \;\; {\rm otherwise}.  \nonumber  
\label{evmess}
\end{eqnarray}
The color constraint on the hopping possibilities is therefore sufficiently 
strong that the result is a diagonal $v$ matrix. The non--uniformity in the
diagonal values is due to the effect of the single $b$ bond, from which one 
color configuration can profit maximally and one cannot profit at all, while 
the other combinations may use this to differing degrees. 

From the manifestly strong effects on the $t$ and $v$ matrices arising 
from the choice of four--site plaquette units, one might conclude that 
symmetry--restoration effects due to summing over all of the available 
plaquettes and orientations are of prime importance, and that in this sense 
the consideration of a QDM does not advance the general understanding beyond 
that obtained in Ref.~[\onlinecite{rno}]. In the multicolored QDM, this 
strong breaking of symmetry will result in a selection process favoring only 
those dimer colors maximizing the contributions from quantum fluctuations, 
and hence in a system of $+$ and $-$ dimers on each active plaquette. 

Whether it is possible to select the preferred dimer colors depends on 
the definition of the system. As in Sec.~III, for models with no dimer 
number--mixing terms and fixed numbers of dimer of each color, the color 
sectors cannot mix and the system is constrained to maximize its energy 
with the available number of colors of each dimer. While the (os/st) case, 
in which the dimer color corresponds to the physical spin, is indeed 
constrained by this type of criterion (below), in the (ss/ot) case it 
is more straightforward to mix bond orbital sectors, as this can be done 
without changing the net electron orbital color balance, and therefore 
a selection of unequal numbers of $T_z$ states is not restricted. Thus 
the effective QDM for the (ss/ot) case is a type of sector--switching 
two--color model with four matrix elements in $v_{ij}$ and one in $t_{ij}$. 

\begin{figure}[t]
\includegraphics[width=4.1cm]{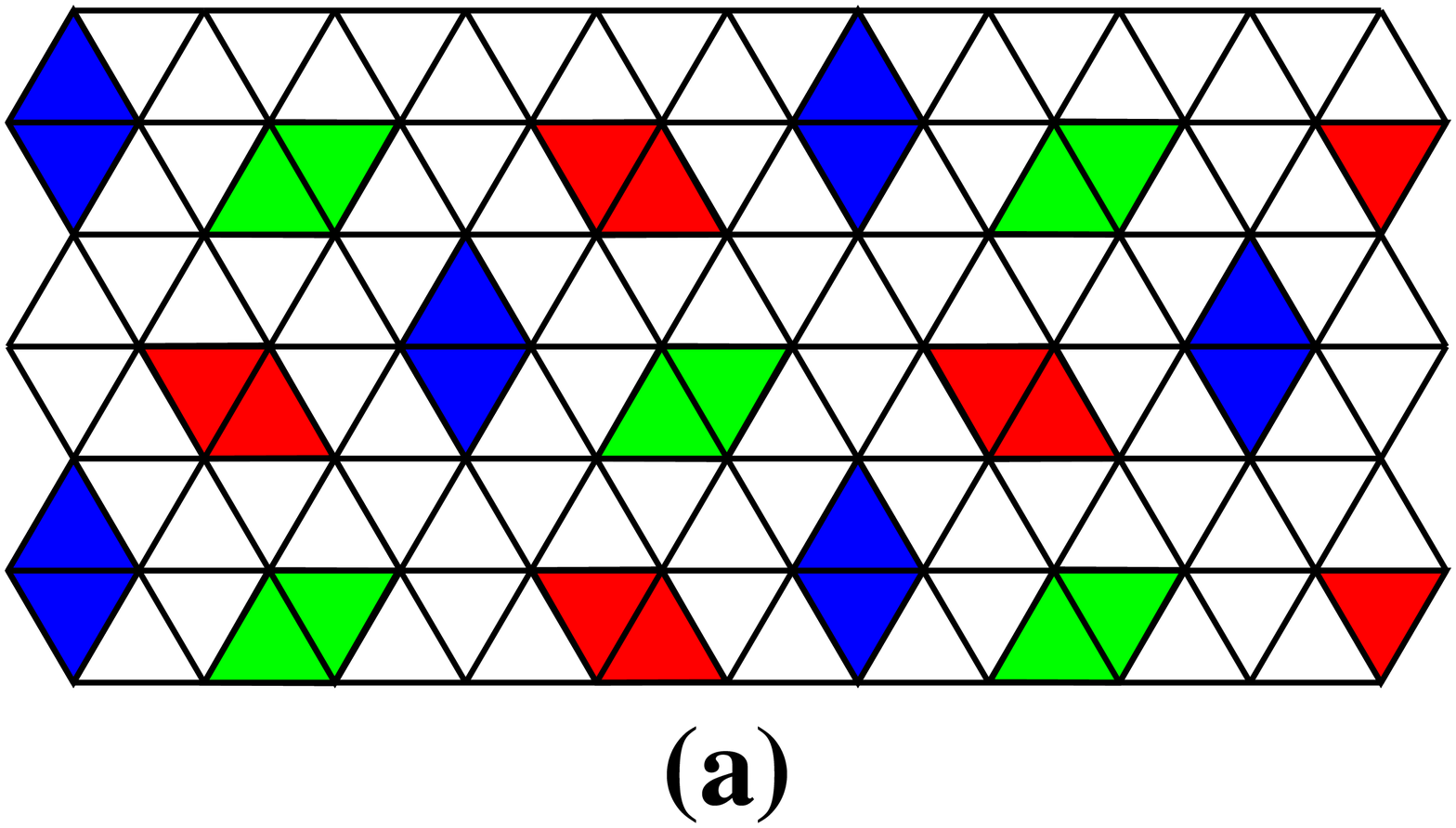}\hspace{0.2cm}\includegraphics[width=4.1cm]{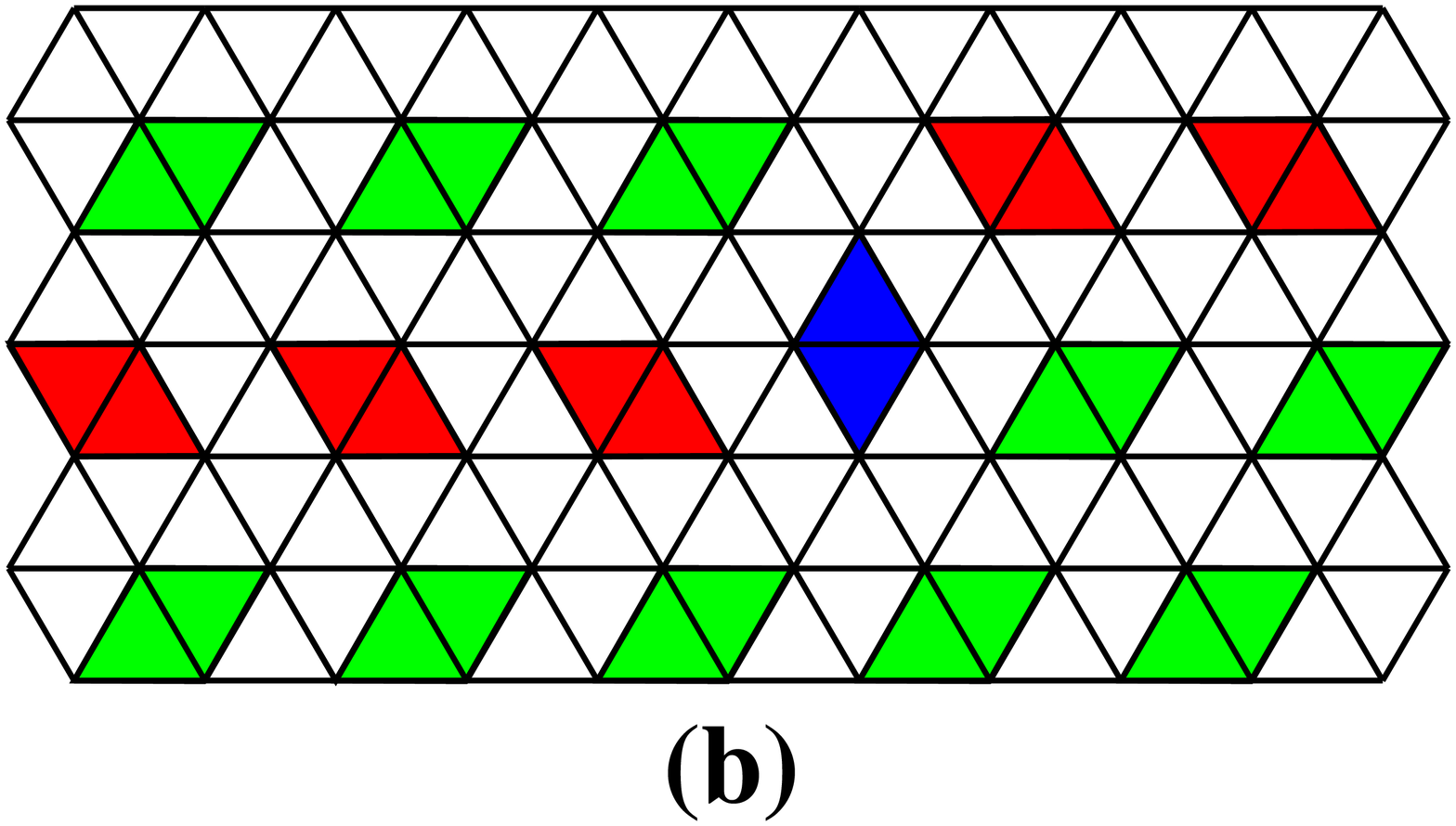} \\
\phantom{.}
\includegraphics[width=4.1cm]{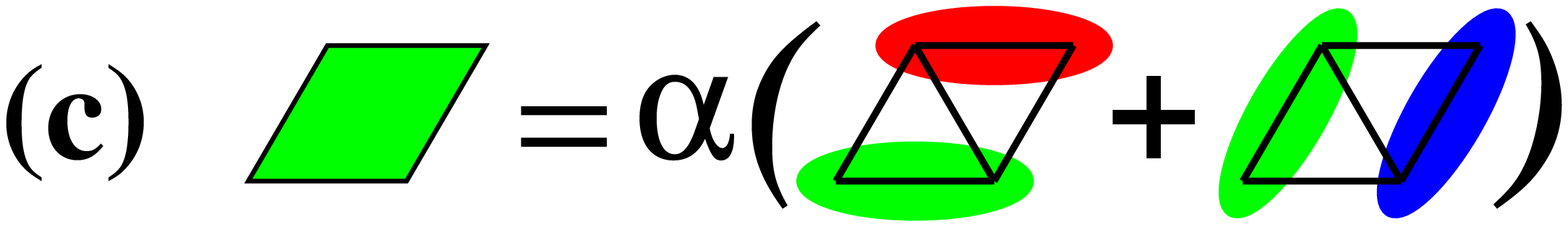}
\caption{(color online) Two plaquette coverings in the ``columnar dimer'' 
class which maximize the number of flippable plaquettes for (ss/ot) dimers 
on the triangular lattice. Shown are (a) a highly regular configuration 
with active plaquettes of all three orientations and (b) a more generic 
configuration of active plaquettes with two orientations plus a single 
``defect'' of the third orientation. The color scheme represents the fact 
(c) that the ground state of each active plaquette is the symmetrized 
combination of states involving dimers with only two out of the possible 
three colors in the relevant orbital triplet; however, this triplet changes 
with the orientation of both the plaquette and the occupied bond. }
\label{qdmf12}
\end{figure}

Such a model, with $|v| > |t|$ and, crucially, $v < 0$, is known to be in 
the ``columnar'' VBC state, the general term applied to the set of dimer 
coverings which maximize the number of flippable plaquettes. Because of 
the constraint that $-+$ plaquettes are inactive for $b$--bond processes, 
the maximal number of active plaquettes obtainable in this ``two--color'' 
model is not the conventional 1/6 but only 1/12. The $v$ and $t$ terms 
select automatically the optimal color state of the spin singlets, and 
the ground state of each active plaquette is the symmetrized combination 
of ``horizontal'' and ``vertical'' dimer pairs with energy $-2J - |v| - t$. 
On the triangular lattice of $N$ sites, the total number of valid plaquette 
states of this type is a highly (but not extensively -- the entropy is 
proportional to the perimeter of the system) degenerate set, of which two 
examples are shown in Fig.~8. That the ground states are VBCs which break 
translational symmetry does not mean that the system has no positional 
fluctuations: rather, this type of configuration maximizes the sole kinetic 
($t$) term present, but this is not strong enough to melt the static 
($v$--driven) plaquette order. In this case, the QDM result is quite 
unambiguous: the ratio deduced for $|t/v|$ is well in the columnar phase, 
and not close to a boundary. 

\subsection{Superexchange limit, physical $\eta$}

Here the ground state of the system is composed of (os/st) spin triplets. 
From the available electronic hopping processes, represented schematically 
in Fig.~9, it is clear that there is no second--order $t$ term which can 
flip the orientation of a pair of (os/st) dimers. In contrast to the (ss/ot) 
case, the calculation of matrix elements thus requires the consideration of 
coherent electronic processes occurring at fourth order in the electronic 
Hamiltonian. For a meaningful comparison between fourth--order $t$ and $v$ 
terms in the relevant QDM, second--order contributions to $v$ of the type 
computed in the previous subsection are treated as a renormalization of 
$J$. Let ${\tilde J}$ denote $4 t^4/(U - 3 J_H)^3$, then 
\begin{eqnarray}
v_{\mu\nu,\mu\nu} & = & - {\tilde J}, \\
v_{\mu\nu,\rho\sigma} & = & 0 \;\; {\rm otherwise}.  \nonumber  
\label{evmeos}
\end{eqnarray}
The spin--conservation symmetry (the absence of a magnetic field is 
assumed) makes the $v$ matrix not only uniform but, in combination with 
the spin--pairing possibilities and the fact that electron hopping involves 
an automatic orbital color change, completely diagonal. The $t$ matrix has 
more variety, because some overlap is permitted within the sectors of fixed 
total spin, and thus 
\begin{eqnarray}
t_{++,++} & = & t_{--,--} \;\; = \;\; {\textstyle \frac{1}{2}} {\tilde J}, 
\nonumber \\
t_{+0,+0} & = & t_{0+,0+} \;\; = \;\; {\textstyle \frac{1}{4}} {\tilde J}, 
\nonumber \\
t_{00,00} & = & {\textstyle \frac{1}{4}} {\tilde J}, \\
t_{0-,0-} & = & t_{-0,-0} \;\; = \;\; {\textstyle \frac{1}{4}} {\tilde J}, 
\nonumber \\
t_{+0,0+} & = & t_{0+,+0} \;\; = \;\; {\textstyle \frac{1}{4}} {\tilde J}, 
\nonumber \\
t_{+-,00} & = & t_{00,+-} \;\; = \;\; {\textstyle \frac{1}{4}} {\tilde J}, 
\nonumber \\
t_{-+,00} & = & t_{00,-+} \;\; = \;\; {\textstyle \frac{1}{4}} {\tilde J}, 
\nonumber \\
t_{-0,0-} & = & t_{0-,-0} \;\; = \;\; {\textstyle \frac{1}{4}} {\tilde J}, 
\nonumber \\
t_{\mu\nu,\rho\sigma} & = & 0 \;\; {\rm otherwise}.  \nonumber  
\label{etmeos}
\end{eqnarray}
In this case, there is blocking of the possible electronic processes leading 
to dimer flips by the allowed spin states: only the all--up and all--down 
spin configurations can profit from every available process. That the element 
$t_{00,00}$ is not symmetrical with $t_{++,++}$ and $t_{--,--}$ is a 
consequence of the breaking of translational symmetry in the plaquette choice. 

\begin{figure}[t]
\includegraphics[width=4.2cm]{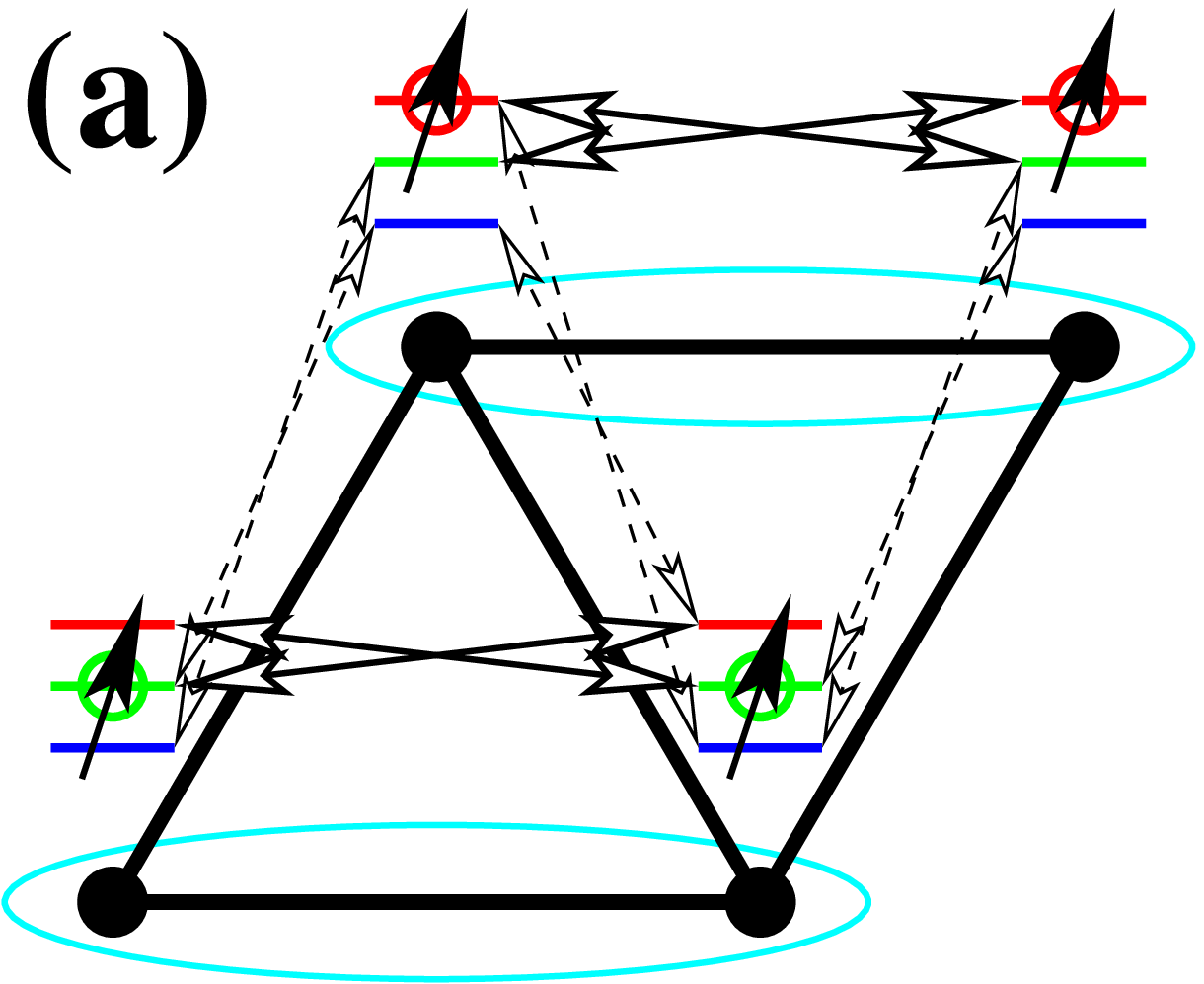}\hspace{0.2cm}\includegraphics[width=4.0cm]{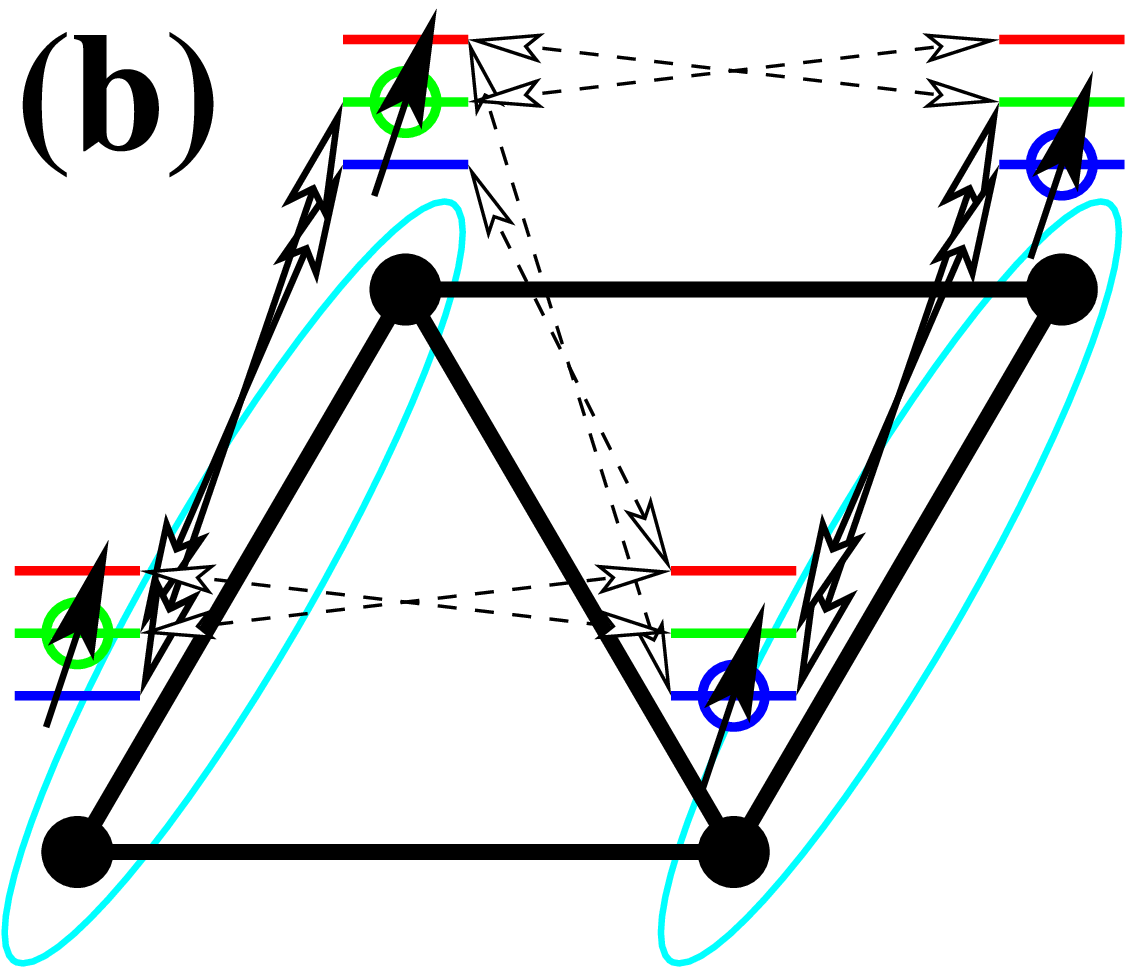}
\caption{(color online) Superexchange limit, $(ac)$ plaquette, showing 
selected (a) horizontal and (b) ``vertical'' (os/st) dimer pairs (turquoise 
ellipses), along with a schematic representation of allowed electron hopping 
processes. Pairs of $S_z = 1$ triplets are depicted in both panels.} 
\label{qdmf13}
\end{figure}

In this case the system is not free to choose its color state, which would 
mean choosing its real magnetic state (in effect this defines another set 
of unmixable ``topological'' sectors characterized by the total $S_z$ of 
the dimer ensemble). Under the reasonable assumption that the ground state 
will be net nonmagnetic, the system will have equal numbers of $+$ and $-$ 
triplets, but the ratio $N_\pm:N_0$ may lie anywhere between 0:1 and 1:0. 
The fact that matrix elements exist for many plaquette processes which mix 
the positions of 0 and $+/-$ triplets suggests that it is most favorable 
for the system to preserve the option of fluctuating among all the possible 
states. If the system were to freeze, locally or globally, to either limiting 
case, the possibilities for dynamical fluctuations would be strongly reduced. 

Thus one has deduced a true, three--color QDM, albeit one with predominantly 
diagonal elements in the 9$\times$9 $t$ and $v$ matrices and with a ratio 
$t_{\mu\mu,\mu\mu}/v_{\mu\mu,\mu\mu} = - {\textstyle \frac{1}{4}}$ for most 
components. This is again a system which is in a robust columnar phase. A 
columnar phase in a system with three dimer colors, all (for the sake of 
illustration) equally occupied, would be one with a one--dimensional nature 
(hard but flexible strings of dimers) of the type shown in Fig.~10. 

Both in this case and for (ss/ot) dimers, individual matrix elements of $v$ 
are rather larger than those in $t$. This result reflects the property 
of the model that all fluctuations away from and back to the starting 
configuration may contribute to $v$, while only quite specific pairs of 
color--switching processes are able to contribute to a $t$ term. This 
places the system in the regime $|v| > |t|$ which is generically unsuitable 
for resonant states, and hence is the dominant physics deciding the (static 
or dynamic) nature of the ground state; the type of static phase is in turn 
determined by the sign of $v$. While it is again necessary to consider the 
restoration of translational symmetries broken rather crudely at the 
four--site plaquette level, it is difficult to conceive of circumstances 
under which this would bridge the gap, which includes changing the sign of 
$v$, from the columnar phase to an RVB phase. 

\begin{figure}[t]
\includegraphics[width=8cm]{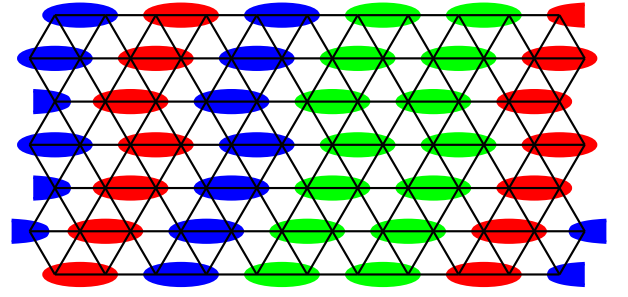}
\caption{(color online) A columnar dimer covering maximizing the number 
of flippable plaquettes on the triangular lattice for (os/st) dimers.}
\label{qdmf14}
\end{figure}

To conclude the analysis of this and the previous subsection, from 
effective QDMs it appears very likely that the $t_{2g}$ superexchange 
models on the triangular lattice favor ``columnar'' VBC states. In the 
(ss/ot) case this is a type of two--color model with plaquette--ordered 
ground states, while in the (os/st) case it is a three--color model with 
dimer--ordered ground states. These states break the symmetry both in real 
space (lattice translation and rotation) and in the space of dimer color. 
This result is a consequence of the fact that virtual fluctuations promoting 
dimer resonance are simply too restrictive (in ``color'', be this the orbital 
type or the real spin) to be numerous enough to compete qualitatively with 
virtual fluctuations profiting from static order. 

\begin{figure}[t]
\includegraphics[width=6.4cm]{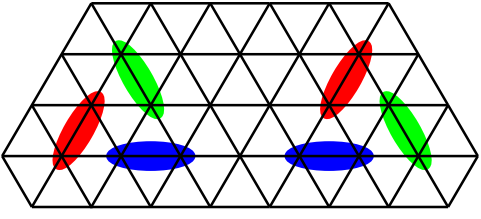}
\caption{(color online) Representation of the six--bond triangular loop 
fluctuation which is the minimal kinetic process permitted in the 
direct--exchange limit.}
\label{qdmf15}
\end{figure}

\subsection{Direct--exchange limit, $\eta = 0$}

As demonstrated in Sec.~II, in this model (\ref{ehd0}) the dimer overlap 
factor is $\alpha = 0$. Thus the model is naturally color--conserving. The 
conventional QDM plaquette terms are simply $t = 0 = v$ and the appropriate 
model is obtained from 6--bond, 3--dimer loops of the type shown in Fig.~3 
(for which there are two orientations). These contribute at sixth order 
in the electronic Hamiltonian, both to a $v'$ term for processes returning 
to the starting configuration and to a $t'$ term for correlated hopping 
processes in the same direction around the triangle, which lead to a net 
shift of the dimer positions. For clarity this is illustrated again in 
Fig.~11, with the dimers retaining their color code (which here, in contrast 
to the preceding subsections, does denote their orbital state). By analogy 
to the procedure mentioned above, fourth--order contributions to maintaining 
the static dimer configuration can also be renormalized into the energy of 
the dimer coverings (whose extensive degeneracy is not affected by these). 
By writing out the wave functions for the triangle states, 
\begin{eqnarray}
|\psi_t^l \rangle & = & {\textstyle \frac{1}{2 \sqrt{2}}} (|1 b \! 
\uparrow 2 b \! \downarrow \rangle - |1 b \! \downarrow 2 b \! \uparrow 
\rangle) (|3 g \! \uparrow 4 g \! \downarrow \rangle \\ & & \;\;\;\; 
 - |3 g \! \downarrow 4 g \! \uparrow \rangle) (|5 r \! \uparrow 6 r \!
\downarrow \rangle - |5 r \! \downarrow 6 r \! \uparrow \rangle), \nonumber \\
|\psi_t^r \rangle & = & {\textstyle \frac{1}{2 \sqrt{2}}} (|2 b \! \uparrow 
3 b \! \downarrow \rangle - |2 b \! \downarrow 3 b \! \uparrow \rangle) 
(|4 g \! \uparrow 5 g \! \downarrow \rangle \\ & & \;\;\;\; 
 - |4 g \! \downarrow 5 g \! \uparrow \rangle) (|6 r \! \uparrow 1 r \! 
\downarrow \rangle - |6 r \! \downarrow 1 r \! \uparrow \rangle), \nonumber 
\label{etwf}
\end{eqnarray}
and proceeding in a manner analogous to the previous two subsections, one 
finds 
\begin{equation}
v' = - 576 \frac{t_e'^6}{U^5}, \;\;\;\; t' \; = \; \frac{603}{2} \frac{t_e'^6}
{U^5}.
\label{edevt}
\end{equation}
As above, the difference in magnitude between $v'$ and $t'$ arises due to the 
prevalence of connected processes not changing the dimer configuration over 
those leading to the specific resonance loop in question. 

Thus once again this is a model strongly in the columnar limit: based on 
a predominant ordering pattern and profiting from limited fluctuations 
(occurring only on selected generalized plaquette units). The meaning 
of the term ``columnar'' in standard QDM literature is a configuration 
maximizing the number of flippable plaquettes, which for the $t'$--$v'$ 
model means the number of up-- and down--oriented triangles. Examples are 
shown in Fig.~12. This columnar ground state has three possible color 
combinations, specifically those where any two colors are used in two 
triangles each, plus a triangle chirality, and hence is six--fold 
degenerate. It does not possess equal numbers of dimers of each color 
(the ratio is 1:1:2). A ground state of a macroscopic system may be 
expected to have a domain structure, and all states with equal dimer 
numbers will present defective versions of this phase. 

The overall energy gain of this state of the QDM relative to the original 
manifold (which has $e_{\rm VB} = - J'/6$) is $e = (v - t) \langle n_{\rm 
fl} \rangle$ per plaquette, which in the present case of six--bond loops 
is $e_6 = (v' - t')/12 \simeq - 73 t_e'^6/U^5$ per bond. However, because 
of the higher--order nature of this correction, it is not able to alter 
the conclusion\cite{rno} that the one--dimensional solution, which has 
energy $e_{\rm 1D} = - {\textstyle \frac{1}{3}} \ln 2 = - 0.23105 J'$ per 
bond, appears to lie lower than all dimer states. It is worth reiterating 
in this context that the present considerations are restricted to the case 
when $\eta$ is identically zero; as noted in Sec.~II, finite values of 
$\eta$ lead to the stabilization of particular static VBC configurations 
which promote energy gains of $O(\eta^3)$,\cite{rji} also insufficient 
to redress the energy balance. The prevalence of the one--dimensional 
solution in a geometry of such high connectivity is testimony to the 
enormous degree of orbital--induced frustration in the full physical system. 

\begin{figure}[t]
\includegraphics[width=4.1cm]{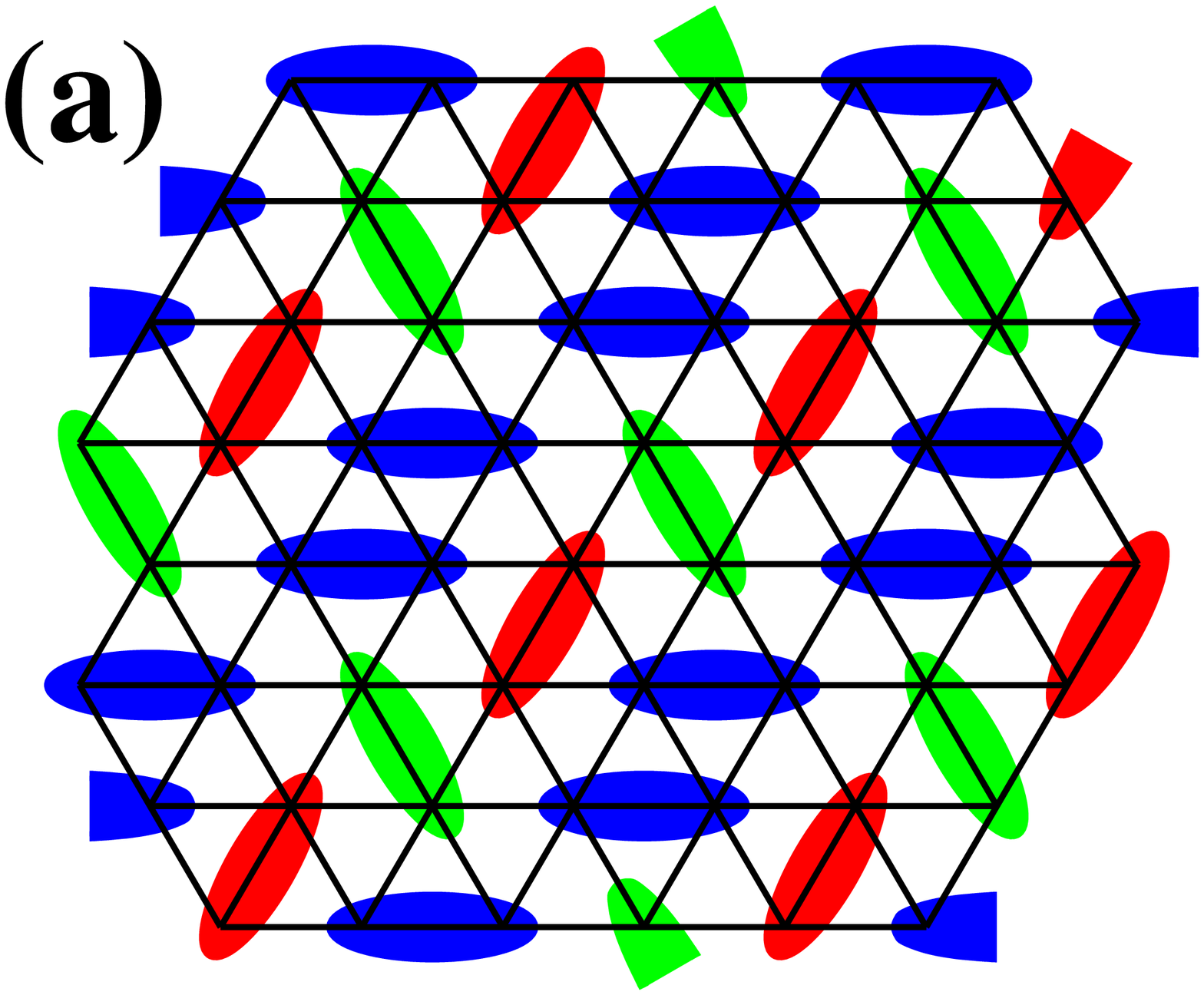}\hspace{0.2cm}\includegraphics[width=4.1cm]{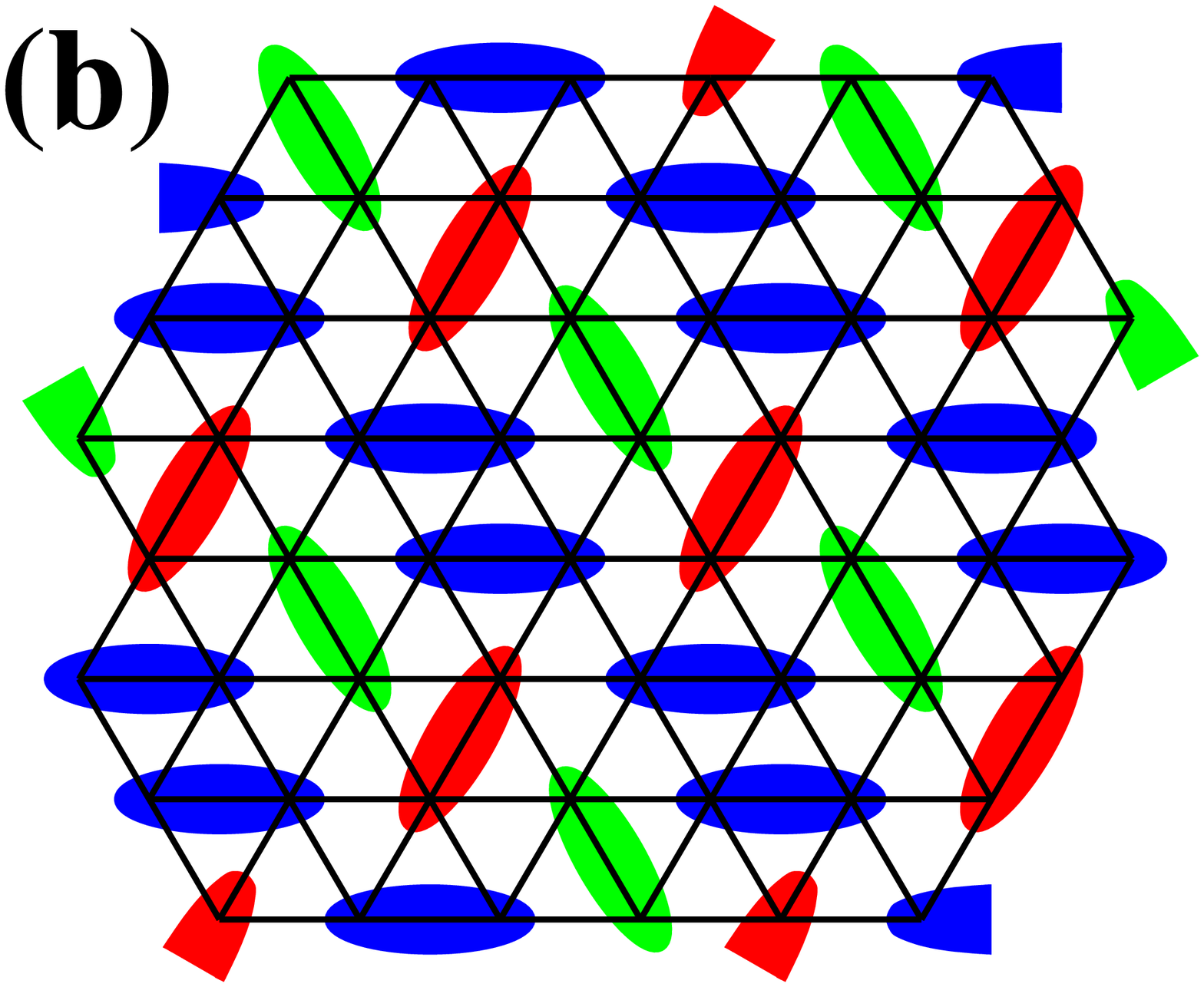}
\caption{(color online) ``Columnar'' dimer coverings of the triangular 
lattice maximizing the number of flippable six--site triangular units. Two 
of the six possibilities are illustrated: in panel (a) the red and green 
dimers (directions) each participate in two loops and the more numerous 
blue dimers in one; in panel (b) is shown the inequivalent state which 
still has blue as the unique direction but has the opposite chirality of 
dimer locations on the the up-- and down--oriented triangular units.}
\label{qdmf16}
\end{figure}

\section{Summary}

This manuscript pursues, by mapping to a set of quantum dimer nodels (QDMs),
the nature of the ground state of the $t_{2g}$ spin--orbital model on the 
triangular lattice. The electronic Hamiltonian, investigated in detail in 
Ref.~[\onlinecite{rno}], represents an insulating $3d^1$ electron system 
on the $\langle 111 \rangle$ planes of a system such as NaTiO$_2$, where 
the cubic structural symmetry of edge--sharing metal--oxygen octahedra
leads to an unbroken, threefold orbital degeneracy. This model was 
shown\cite{rno} to have a very strong preference for dimer--based states 
of no long--ranged magnetic or orbital order over its entire parameter 
range in both the ratios $J'/J$ (of the direct and superexchange contributions 
to the magnetic interactions) and $J_H/U$ (of the Hund coupling to the 
on--site Coulomb repulsion). 

These dimer states were found to span a range of behavior from highly resonant 
in the superexchange limit to quasi--static in the direct--exchange limit. 
Both cases present fundamental challenges in determining the nature of the 
true ground state: the former is a candidate resonating valence--bond (RVB) 
state and the latter a situation where subtle selection effects, sometimes 
known as ``order--by--disorder,'' are responsible for choosing the true ground 
state from a very highly degenerate manifold. In Ref.~[\onlinecite{rno}] only 
energetic studies were performed, which were unable to answer the topological 
questions associated with the formation of an RVB state or to resolve the 
possible differences among quasi--static valence--bond coverings.

These questions are addressed by mapping the electronic Hamiltonian to the 
minimal QDM for each of three cases: (a) the pure superexchange limit for 
low $J_H/U$, (b) the pure superexchange limit for intermediate $J_H/U$, and 
(c) the direct--exchange limit for $J_H/U = 0$. In case (a), the ground state 
is composed of spin--singlet, orbital--triplet (ss/ot) dimer entities and 
the effective QDM has three dimer ``colors'' corresponding to their orbital 
state. In case (b), the ground state is composed of orbital--singlet, 
spin--triplet (os/st) dimers and the effective QDM has again three colors, 
which correspond to the triplet spin components of the dimers. In case (c), 
there are three dimer colors corresponding to the orbital colors active in 
each of the three bond directions of the triangular lattice, but because 
these are locked to each other, the color is not a degree of freedom and 
the effective QDM has one color. The ``non--orthogonality catastrophe'' 
which affects conventional spin dimers, namely that all dimer coverings 
have finite overlap, is shown to be very strongly reduced or completely 
eliminated in the presence of a dimer color in such models. 

In order to analyze the properties of these specific QDMs, first the general 
multicolored QDM must be understood. This is investigated for the two--color 
case, with some additional consideration of three--color QDMs where these 
contain further physics. Both the potential term $v$ for adjacent dimers and 
the kinetic term $t$ for these to flip direction on the four--site plaquette 
they define become $n^2\!\times\!n^2$ matrices, where $n$ is the number of 
colors. With such a choice of matrix elements, there remains only one 
Rokhsar--Kivelson (RK) point, where the equally weighted superposition of 
all possible dimer coverings (in a given topological sector) is an eigenstate, 
but the regime of parameters in which this (gapped, RVB) state is the ground 
state is expanded considerably. Non--trivial topological sectors are found 
for specific choices of matrix elements, notably those conserving the total 
number of dimers of each color and those which allow this to be altered not 
in single steps but in units of two. This gives rise to topological 
excitations related to the color degree of freedom, which are termed 
``color visons.'' 

The matrix elements $t_{\mu\nu,\rho\sigma}$ and $v_{\mu\nu,\rho\sigma}$ 
are deduced from the electronic Hamiltonian for cases (a) and (b). The 
restrictions on hopping of electrons of different orbital color and the 
breaking of translational symmetry contained in the consideration of 
four--site plaquettes lead in case (a) to very sparse and asymmetric 
matrices. This asymmetry selects only the $T_z = \pm 1$ orbital triplet 
states, and the model is reduced to one with two colors. In case (b), the 
physical spin defines ``topological'' sectors which cannot mix. In both 
cases, elements of the $v$ matrix are generally rather larger than those 
of the $t$ matrix, and have in addition a negative sign. Thus the mapping 
to a QDM indicates strongly that the ground state of both models is a 
``columnar'' plaquette phase, one which is based on a small subset of 
static dimer coverings which maximize the number of flippable four--site 
plaquettes: this number is 1/12 of the total in case (a) and 1/6 in case 
(b). While these states gain energy from virtual dimer flipping processes 
(quantum fluctuations), such processes are not sufficiently strong that 
they can ``melt'' the plaquette or dimer order in favor of a completely 
resonant, symmetry--restored phase.

In case (c), the overlap matrix elements, and $t$ and $v$ elements, on 
a four--site plaquette are identically zero. The leading quantum fluctuation 
processes for dimer resonance occur on six--site, triangular units containing 
one dimer of each bond direction, and hence the corresponding minimal QDM is 
a $t'$--$v'$ model defined on these ``plaquettes''. Once again, $v'$ is 
negative and larger in magnitude than $t'$, suggesting that the ground 
state of the system is again a generalized columnar covering (meaning one 
which maximizes the number of flippable triangles). These are illustrated 
and shown to have a six--fold degeneracy, corresponding to the bond direction 
and the triangle chirality. 

In summary, effective QDMs are used to obtain indications as to the nature of 
the ground state of a complex electronic Hamiltonian for a spin--orbital model 
which is known to have no conventional magnetic or orbital order. In all cases 
of most interest, the ground state is found to be a resonance--stabilized 
spin--orbital VBC covering which maximizes quantum fluctuation processes on 
a restricted set of plaquettes, but does not allow the melting of a preferred 
order by these fluctuations. While the derivation of a QDM involves rather 
crude symmetry--breaking at the level of four-- (or six--)site plaquettes, 
and the symmetry--restoring effects of fluctuations across all available 
plaquettes must be borne in mind, the columnar phases deduced from all 
three QDMs are quite robust, indicating that there is no nearby liquid 
phase.

\acknowledgments 
 
The author is especially grateful for the invaluable contributions of G. 
Jackeli. Thanks are due also to D. Ivanov and F. Mila for helpful discussions. 
This work was supported by the National Science Foundation of China under 
Grant No.~10874244 and by Chinese National Basic Research Project 
No.~2007CB925001.


\begin{thebibliography}{99} 

\bibitem{rno} B. Normand and A. M. Ole\'s, Phys. Rev. B {\bf 78}, 094427 
(2008).

\bibitem{rmr} see R. Moessner and K. S. Raman, in {\sl Highly Frustrated 
Magnetism}, eds. C. Lacroix, P. Mendels, and F. Mila (Springer, Heidelberg, 
2010), and references therein. 

\bibitem{rmvrbfp} F. Mila, F. Vernay, A. Ralko, F. Becca, P. Fazekas, and 
K. Penc, J. Phys.: Condens. Matter {\bf 19}, 145201 (2007).

\bibitem{rrk} D. S. Rokhsar and S. A. Kivelson, Phys. Rev. Lett {\bf 61}, 
2376 (1988).

\bibitem{rms} R. Moessner and S. L. Sondhi, Phys. Rev. Lett {\bf 86}, 1881 
(2001).

\bibitem{rvrbm} F. Vernay, A. Ralko, F. Becca, and F. Mila, Phys. Rev. B 
{\bf 74}, 054402 (2006). 

\bibitem{rcbs} G. Chen, L. Balents, and A. P. Schnyder, Phys. Rev. Lett. 
{\bf 102}, 096406 (2009); G. Chen, A. P. Schnyder, and L. Balents, Phys. 
Rev. B {\bf 80}, 224409 (2009). 

\bibitem{rkk} K. I. Kugel and D. I. Khomskii, Usp. Fiz. Nauk {\bf 136}, 
621 (1982) [Sov. Phys. Usp. {\bf 25}, 231 (1982)].

\bibitem{rji} G. Jackeli and D. A. Ivanov, Phys. Rev. B {\bf 76}, 132407 
(2007).

\bibitem{rw} X.--G. Wen, {\sl Quantum Field Theory of Many-Body Systems}, 
(OUP, Oxford, 2004).

\bibitem{rrfbim} A. Ralko, M. Ferrero, F. Becca, D. Ivanov, and F. Mila,
Phys. Rev. B {\bf 71}, 224109 (2005). 

\bibitem{riif} A. Ioselevich, D. A. Ivanov, and M. V. Feigelman, Phys. Rev. 
B {\bf 66}, 174405 (2002).

\bibitem{rk} P. W. Kasteleyn, J. Math Phys. {\bf 4}, 287 (1963); S. Samuel, 
J. Math Phys. {\bf 21}, 2806 (1980). 

\end{thebibliography}
\end{document}